\documentclass[aps,pre,preprint,superscriptaddress,compress]{revtex4-2}

\usepackage{booktabs}
\usepackage{threeparttable}
\usepackage{graphicx}
\usepackage{dcolumn}
\usepackage{bm}
\usepackage{epsfig}
\usepackage{booktabs}
\usepackage{subfigure}
\usepackage{graphics}
\usepackage{amssymb}
\usepackage{amsmath}
\usepackage{array}
\usepackage{color}
\usepackage{ulem}
\usepackage{commath}
\usepackage{mathtools}
\usepackage{txfonts}
\usepackage{mathtools}
\usepackage{multirow}
\usepackage{CJKutf8}

\usepackage{lineno,hyperref}
\graphicspath{{./pic/}}

\usepackage{float}
\begin{document}
	\begin{CJK*}{UTF8}{gbsn}
		\preprint{APS}
		
		\title{Electrohydrodynamics of a pair of leaky dielectric droplets on the solid substrate: A lattice Boltzmann study}
		
		
		
		\author{Jiang Peng}
		\affiliation{School of Mathematics and Statistics, Huazhong University of Science and Technology, Wuhan, 430074, China}
		\author{Xi Liu}
		\email[Corresponding author: ]{aubrey_xixi@126.com}
		\affiliation{School of Mathematics and Statistics, Huazhong University of Science and Technology, Wuhan, 430074, China}
		\affiliation{Institute of Interdisciplinary Research for Mathematics and Applied Science, Huazhong University of Science and Technology, Wuhan 430074, China}
		\affiliation{Hubei Key Laboratory of Engineering Modeling and Scientific Computing, Huazhong University of Science and Technology, Wuhan, 430074, China}
		\author{Zhenhua Chai}
		\email[Corresponding author: ]{hustczh@hust.edu.cn}
		\affiliation{School of Mathematics and Statistics, Huazhong University of Science and Technology, Wuhan, 430074, China}
		\affiliation{Institute of Interdisciplinary Research for Mathematics and Applied Science, Huazhong University of Science and Technology, Wuhan 430074, China}
		\affiliation{Hubei Key Laboratory of Engineering Modeling and Scientific Computing, Huazhong University of Science and Technology, Wuhan, 430074, China}
		\author{Changsheng Huang}
		\affiliation{School of Mathematics and Statistics, Huazhong University of Science and Technology, Wuhan, 430074, China}
		\affiliation{Institute of Interdisciplinary Research for Mathematics and Applied Science, Huazhong University of Science and Technology, Wuhan 430074, China}
		\affiliation{Hubei Key Laboratory of Engineering Modeling and Scientific Computing, Huazhong University of Science and Technology, Wuhan, 430074, China}
		\author{Xiufang Chen}
		\affiliation{School of Mathematics and Statistics, Huazhong University of Science and Technology, Wuhan, 430074, China}
		
		
		
		\date{\today}
		
		\begin{abstract}
			In this work, the electrohydrodynamics of a pair of leaky dielectric droplets on a solid substrate is investigated by the phase-field-based lattice Boltzmann method. Different from a pair of suspended droplets that may coalesce or separate, two leaky dielectric droplets on the substrate exhibit more complex modes due to the effects of wettability and electric force. The results show that when a horizontal electric field is applied, five different modes with electrostatic attractive force are observed, including attraction without coalescence, attraction with coalescence, coalescence with bubble entrapment, coalescence followed by suspension, and suspension followed by coalescence. Particularly, if the droplets are in a hydrophilic state, the coalescence mode is usually observed, while for droplets in a neutral or hydrophobic state, the permittivity ratio $S$ has an important effect on the droplet modes. Additionally, during the coalescence process, two droplets in a hydrophobic state not only capture bubbles, but may also exhibit suspension at a large permittivity ratio $S$ or contact angle $\theta$. On the other hand, when a vertical electric field is applied, there are three different modes with repulsive electrostatic force, including non-coalescence, coalescence, and suspension followed by repulsion. Specially, a small permittivity ratio $S$ or a large contact angle $\theta$ can suppress the horizontal deformation of droplets, preventing their coalescence. 	Moreover, under superhydrophobic conditions, both horizontal and vertical electric fields suspend the droplets. However, the vertical electric field induces repulsion between the suspended droplets, driving them apart, whereas the horizontal electric field promotes their coalescence.
		\end{abstract}


\maketitle
\end{CJK*}
\section{\label{sec:introduc.}Introduction}
Controlling droplet coalescence is a significant challenge in droplet microfluidics. On one hand, to facilitate chemical reactions through mixing, promoting droplet coalescence is essential for merging specific droplets \cite{Ahn2006}. On the other hand, to manage a large number of droplets, it is desirable to inhibit coalescence to maintain the intact and distinct partitioning of each droplet \cite{Hartmann2022}. As one of the manipulation technologies, the adoption of an external electric field can be used to control droplet deformation, breakup, coalescence, or repulsion, thereby fulfilling the stringent demands of both scientific research and industrial applications \cite{An2015}.

In recent years, many theoretical, experimental and numerical studies have been performed to investigate the interaction of two droplets under an external electric field \cite{Sozou1975,Atten1993,Atten2006,Zhang1995,Dong2002,Bird2009,Huang2020,Eow2003,Baygents1998,Dong2018,Das2021,Sorgentone2021}. Sozou \cite{Sozou1975} first theoretically studied a droplet pair aligned along the electric field direction, where the leaky dielectric droplets are suspended in a leaky dielectric fluid (both droplets and fluid have finite conductivities). It was found that when the distance between the centers of two droplets is reduced to the order of twice droplet diameter, the tangential electric stress at the interfaces of droplets and shear stress induced by the flow field are significantly affected by each other. Then, Atten \cite{Atten1993} estimated the coalescence rate and the characteristic coalescence time of two emulsion droplets via considering the interaction of dipoles induced by the electric field. Moreover, Atten et al. \cite{Atten2006} theoretically derived an approximate correlation between the critical coalescence distance and electric field strength when investigating electrocoalescence conditions of two droplets suspended in an insulating oil. In addition, by taking into account both hydrodynamic and electrostatic interactions, Zhang et al. \cite{Zhang1995} utilized the trajectory analysis to track the relative motion of two perfectly conducting droplets, and predicted their collision and coalescence rates. However, due to the perfect dielectric assumption, these theoretical results cannot predict the dynamic behavior of leaky dielectric droplets. To overcome the limitations of theoretical approaches, experimental investigations have been undertaken to explore the complex behaviors of droplet pairs under electric fields. For instance, Dong et al. \cite{Dong2002} studied the effect of electric field on the coalescence of two droplets suspended in the organic fluid, and demonstrated that the reduced surface tension promotes droplet deformation, facilitating the contact coalescence of two droplets. They also observed that coalescence time decreases with increasing electric field strength. Bird et al. \cite{Bird2009} observed that the leading edges of two close droplets can deform into cones due to the electrical stress acting on the droplet interfaces, and whether they coalesce after contact depends on the cone angle. Recently, Huang et al. \cite{Huang2020} studied the dynamic behaviors of two free droplets suspended in low-viscosity oil under a uniform electric field. They found that as the electric field strength increases, the two droplets undergo a transition from coalescence to partial coalescence and then to non-coalescence, and the critical value of electric field strength for droplet coalescence is related to the droplet radius and surface tension. Furthermore, Eow and Ghadiri \cite{Eow2003} experimentally investigated the interaction between a pair of suspended droplets, and found that the angle $\alpha$ between the electric field direction and the line connecting the centers of two droplets has an influence on the attractive or repulsive interaction between the two droplets. Although experimental methods offer a distinct advantage in directly observing and measuring the electrohydrodynamics of a droplet pair under an external electric field, they also suffer from some significant limitations in providing a complete and detailed description of the characteristics of flow and electric fields. Consequently, with advances in computational science, numerical approaches have emerged as powerful alternatives. Early numerical work by  Baygents et al. \cite{Baygents1998} studied the interaction between the leaky dielectric droplets aligned in the direction of electric field. They found that there are some circulation flows driven by the electric stress on the droplet surface, whether the droplets attract or repel each other. Later, Dong et al. \cite{Dong2018} investigated the dynamic behavior of the leaky dielectric droplets by using the lattice Boltzmann (LB) method, and the results show that when the permittivity ratio $S$ between the droplet and the surrounding fluid is much smaller than the conductivity ratio $R$, the droplets repel each other. This repulsion is primarily due to the dominant repulsive interaction under a small $S$. On the contrary, when $S$ is large but still smaller than $R$, the droplets can attract each other through the attractive electrostatic interaction. For the case where $S$ is larger than $R$, the droplets only attract each other, and then coalesce, resulting from the combined effects of both hydrodynamic and electrostatic interactions. It is worth noting that an estimator to measure the relative strength of electrostatic force to hydrodynamic force has been proposed \cite{Das2021}, which indicates that the interaction between two suspended droplets cannot be determined only by the sign of $(R-S)$, but also by the relative strength of above two forces and the inertial effect characterized by the Ohnesorge number ($Oh$). Here, $Oh$ also represents the ratio of the viscous capillary to Rayleigh inertial time scale, serving as a metric to assess the influence of viscosity on interfacial dynamics. Furthermore, the inclination angle $\alpha$ also has a significant effect on the electrohydrodynamics of two leaky dielectric droplets through the dielectrophoretic (DEP) and electrohydrodynamic (EHD) interactions \cite{Sorgentone2021}. Specifically, the DEP interaction exerts either an attractive or repulsive influence that is contingent upon the angle $\alpha$,  whereas the EHD interaction is jointly determined by the angle $\alpha$ and the sign of $(R-S)$. 

The aforementioned works have indeed promoted the understanding of the interaction between a pair of droplets under an external electric field. However, when droplets are manipulated under an electric field, they are inevitably in contact with the solid substrate. In this case, the wettability of the solid substrate would affect the equilibrium shapes and dynamic behaviors of the droplets, but this effect has not been comprehensively studied. To fill the gap, in this work we will conduct a numerical study on the electrohydrodynamics of a droplet pair on the solid substrate. For this purpose, the phase-field lattice Boltzmann (LB) method \cite{Kruger2017} is adopted here owing to its distinct features in the simulation of multiphase EHD flows \cite{Liu2019,Liu2021_cicp,Liu2024,Hu2025}, including a clear physical picture, easy implementation of boundary conditions, and fully parallel algorithm \cite{Chen1998,Zhan2022,Liu2025}. The rest of this paper is organized as follows. In section \ref{sec2}, we first present the mathematical model for the physical problem of a pair of leaky dielectric droplets on the solid substrate, followed by the phase-field LB method. Then in section \ref{validation}, the LB method is tested by two classical multiphase examples. In section \ref{Results}, the effects of the permittivity ratio $S$ and wettability on the droplet pair are investigated when the electric field is applied in horizontal or vertical direction. Finally, some conclusions are given in section \ref{sec5}.

\section{Mathematical model and numerical method}\label{sec2}
\subsection{The consistent and conservative phase-field-based model for two-phase EHD flows}
In this work, we only consider the electrohydrodynamics of two droplets that have no net charge but finite electrical conductivity. In the framework of the leaky dielectric theory, a conservative Allen-Cahn equation is used to capture the fluid-fluid interface, the incompressible Navier-Stokes equations are adopted to describe the fluid motion, and the Poisson equation is applied to depict the distribution of electric potential \cite{Hua2008,Liu2019}: 
\begin{equation}
	\frac{\partial C}{\partial t}+\nabla \cdot(C \mathbf{u})=\nabla \cdot[M(\nabla C-\lambda \mathbf{n})],
	\label{phasefield}
\end{equation}
\begin{subequations}
	\begin{equation}
		\nabla \cdot \mathbf{u}=0,
		\label{ns1}
	\end{equation}
	\begin{equation}
		\frac{\partial (\rho \mathbf{u})}{\partial t}+\nabla \cdot(\rho \mathbf{u u}+\mathbf{m}^C\mathbf{u})=-\nabla p+\nabla \cdot\mu\left[\nabla \mathbf{u}+(\nabla \mathbf{u})^{\mathrm{T}}\right]+\mathbf{F},
		\label{ns2}
	\end{equation}
	\label{ns}
\end{subequations}
\begin{equation}
	\nabla \cdot(\sigma \nabla \phi)=0,
	\label{poisson}
\end{equation}
where $C$ is the order parameter used to label the liquid phase (marked as $l$) with $C_l$ = 1, and gas or another liquid phase (marked as $g$) with $C_g$ = 0. $\mathbf{u}$ is the fluid velocity, $M$ is the mobility, $\mathbf{n}=\nabla C/\left|\nabla C\right|$ is the unit vector normal to interface, and $\lambda=\sqrt{2\beta/\kappa}(C_l-C)(C-C_g)$ with $\beta=12\gamma/[W(C_l-C_g)^4]$ and $\kappa=3\gamma W/[2(C_l-C_g)^2]$, $\beta$ and $k$ are two physical parameters related to the interface thickness $W$ and the surface tension coefficient $\gamma$. The mass diffusion $\mathbf{m}^C$ between different phases can be expressed as
\begin{equation}
	\mathbf{m}^C=-\frac{\rho_l-\rho_g}{C_l-C_g}M(\nabla C-\lambda\mathbf{n}),
\end{equation}
which can be used to preserve the mass and momentum consistency \cite{Zhan2022}. $\rho$ is the density, $p$ is the pressure, $\mu$ is the dynamic viscosity, and $\mathbf{F}$ is the total external force. $\sigma$ is the conductivity, $\phi$ is the electrical potential. Under the assumption of $\nabla \times\mathbf{E}=0$, the electric field strength $\mathbf{E}$ can be expressed as a gradient of the electrical potential, i.e., $\mathbf{E}=-\nabla\phi$. 

When the gravity is neglected, the total external force $\mathbf{F}$ is composed of the surface tension force $\mathbf{F}_S$ and electric stress $\mathbf{F}_E$, which can be given by
\begin{equation}
	\mathbf{F}_S=\mu_C\nabla C,\quad	\mathbf{F}_E=-\frac{1}{2}\mathbf{E} \cdot \rho_e\nabla \varepsilon + \rho_e \mathbf{E},
\end{equation}
where $\mu_C$ is the chemical potential defined by
$\mu_C=4\beta(C-C_l)(C-C_g)(C-\frac{C_l+C_g}{2})-k\nabla^2 C$, $\varepsilon$ is the permittivity, and $\rho_e=\nabla \cdot (\varepsilon\mathbf{E})$ is the volume charge density. In order to simplify and improve the accuracy in computing electric stress $\mathbf{F}_E$, we can rewrite the volumetric charge density $\rho_e$ as
\begin{equation}
	\rho_e=\nabla \cdot (\varepsilon\mathbf{E})=\varepsilon \nabla \cdot \mathbf{E}+\nabla \varepsilon \cdot \mathbf{E}=-\frac{\varepsilon}{\sigma}\nabla \sigma \cdot \mathbf{E}+\nabla \varepsilon \cdot \mathbf{E},
\end{equation}
where Eq. (\ref{poisson}) has been used.

To ensure that the distribution of each physical quantity or parameter $\chi$ in the two-phase system is continuous, the following relation is adopted \cite{He1999},
\begin{equation}
	\chi=\frac{\chi_l-\chi_g}{C_l-C_g}(C-C_g)+\chi_g,
\end{equation}
where $\chi$ can be the fluid density $\rho$, the dynamic viscosity $\mu$, the conductivity $\sigma$ or the permittivity $\varepsilon$.

\subsection{Boundary conditions}
\begin{figure}[h]
	\centering
	\includegraphics[width=4.0in]{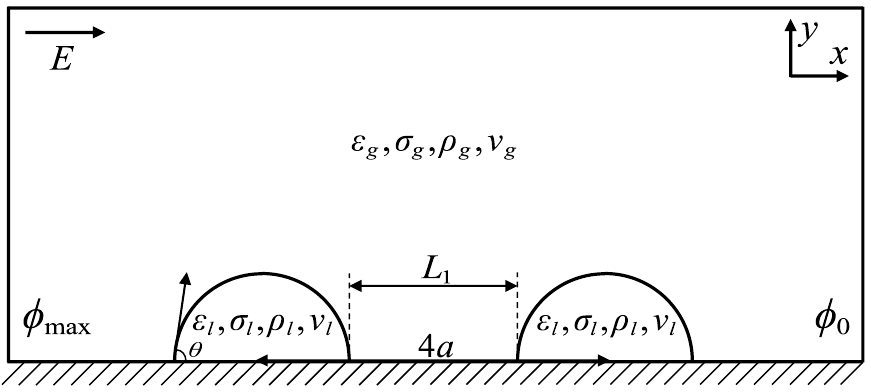}
	\caption{Schematic of a droplet pair placed on the solid substrate under an external electric field.}
	\label{picture-gkt_xm}
\end{figure}

As shown in Fig. \ref{picture-gkt_xm}, when the droplets contact with the solid substrate, the wettability must be considered. In this work, the following wetting boundary condition is used \cite{Huang2015,Liang2019},
\begin{equation}
	\mathbf{n}_w \cdot \nabla C=-\sqrt{\frac{2\beta}{k}}(C_l-C)(C-C_g)\mathrm{cos}\theta,
\end{equation}
where $\mathbf{n}_w$ denotes the unit vector normal to the solid substrate, and $\theta$ represents the contact angle. In addition, for the electric and flow fields, the Dirichlet boundary conditions and the no-slip boundary conditions are imposed on the electrode plates, respectively.

\subsection{The consistent and conservative phase-field-based LB method}
\subsubsection{LB models for the Allen-Cahn equation, Navier-Stokes equations and Poisson equation}
\label{LB model}
Compared to the popular and simple single-relaxation-time LB model \cite{Qian1992}, here we consider the multiple-relaxation-time (MRT) LB model for its better numerical accuracy and stability \cite{Chai2020}. The evolution equations of the MRT-LB model for Allen-Cahn equation [Eq. (\ref{phasefield})], Navier-Stokes equations [Eq. (\ref{ns})] and Poisson equation [Eq. (\ref{poisson})] can be given by \cite{Zhan2022}
\begin{subequations}
	\begin{equation}
		f_i\left(\mathbf{x}+\mathbf{c}_i \Delta t, t+\Delta t\right)-f_i(\mathbf{x}, t)=-\mathbf{\Lambda}^{f}_{ij} \left[f_j(\mathbf{x}, t)-f_j^{e q}(\mathbf{x}, t)\right]+\Delta t(\delta_{ij}-\frac{\mathbf{\Lambda}^{f}_{ij}}{2})F_j(\mathbf{x},t),
		\label{evolution_phasefield}
	\end{equation}
	\begin{equation}
		g_i\left(\mathbf{x}+\mathbf{c}_i \Delta t, t+\Delta t\right)-g_i(\mathbf{x}, t)=-\mathbf{\Lambda}_{ij}^{g} \left[g_j(\mathbf{x}, t)-g_j^{e q}(\mathbf{x}, t)\right]+\Delta t(\delta_{ij}-\frac{\mathbf{\Lambda}_{ij}^{g}}{2})G_j(\mathbf{x},t),
		\label{evolution_ns}
	\end{equation}
	\begin{equation}
		h_i\left(\mathbf{x}+\mathbf{c}_i \Delta t, t'+\Delta t\right)-h_i(\mathbf{x}, t')=-\mathbf{\Lambda}_{ij}^{h} \left[h_j(\mathbf{x}, t')-h_j^{e q}(\mathbf{x}, t')\right],
		\label{evolution_poisson}
	\end{equation}
\end{subequations}
where $f_i(\mathbf{x},t)$ and $g_i(\mathbf{x},t)$ are the distribution functions for phase and flow fields at position $\mathbf{x}$ and time $t$, $h_i(\mathbf{x},t')$ is the distribution function for electric field at position $\mathbf{x}$ and pseudo time $t'$. $\mathbf{c}_i$ is the discrete velocity, $\mathbf{\Lambda}^{f}_{ij}=\left(\mathbf{M}_{f}^{-1} \mathbf{S}_f \mathbf{M}_{f}\right)_{i j}$, $\mathbf{\Lambda}^{g}_{ij}=\left(\mathbf{M}_{g}^{-1} \mathbf{S}_g \mathbf{M}_{g}\right)_{i j}$ and $\mathbf{\Lambda}^{h}_{ij}=\left(\mathbf{M}_{h}^{-1} \mathbf{S}_h \mathbf{M}_{h}\right)_{i j}$ are the collision matrices. Here $\mathbf{M}_f$, $\mathbf{M}_g$ and $\mathbf{M}_h$ are the orthogonal transformation matrices, $\mathbf{S}_f$, $\mathbf{S}_g$ and $\mathbf{S}_h$ are the diagonal relaxation matrices. To recover the Allen-Cahn equation, Navier-Stokes equations and Poisson equation correctly, the equilibrium distribution functions $f_i^{eq}(\mathbf{x},t)$, $g_i^{eq}(\mathbf{x},t)$ and $h_i^{eq}(\mathbf{x},t')$ can be designed as follows \cite{Chai2016,Ren2016,Guo2000,Chai2008}
\begin{subequations}
	\begin{equation}
		f_i^{eq}(\mathbf{x},t)=\omega_i C \big( 1+\frac{\mathbf{c}_i \cdot \mathbf{u}}{c_s^2} \big),
	\end{equation}
	\begin{equation}
		g_i^{eq}(\mathbf{x},t)= 
		\begin{cases} \rho_0 + 
			(\omega_0-1)\frac{p}{c_s^2}+s_0(\rho,C,\mathbf{u}), & i=0,\\ 
			\omega_i\frac{p}{c_s^2}+s_i(\rho,C,\mathbf{u}), & i\not=0,
		\end{cases}
	\end{equation}
	\begin{equation}
		h_i^{eq}(\mathbf{x},t')= 
		\begin{cases} 
			(\omega_0-1)\phi(\mathbf{x},t'), & i=0,\\ 
			\omega_i\phi(\mathbf{x},t'), & i\not=0,
		\end{cases}
	\end{equation}
\end{subequations}
where $\rho_0=1$, $c_s$ is the sound speed, $\omega_i$ is the weight coefficient, and $s_i(\rho,C,\mathbf{u})$ can be given by 
\begin{equation}
	s_i(\rho,C,\mathbf{u})=\omega_i \big[\frac{\mathbf{c}_{i} \cdot (\rho\mathbf{u})}{c_s^2}+\frac{(\rho\mathbf{u}\mathbf{u}+\mathbf{m}^C\mathbf{u}):(\mathbf{c}_{i}\mathbf{c}_{i}-c_s^2\mathbf{I})}{2c_s^4} \big].
\end{equation}
The source and force distribution functions $F_i(\mathbf{x},t)$ and $G_i(\mathbf{x},t)$ can be expressed as \cite{Chai2016,Ren2016,Zhan2022}
\begin{subequations}
	\begin{equation}
		F_i(\mathbf{x},t)=\frac{\omega_i \mathbf{c}_i \cdot [\partial_t (C \mathbf{u}) + c_s^2 \lambda \mathbf{n}]}{c_s^2},	
	\end{equation}
	\begin{equation}
		G_i(\mathbf{x},t)=\omega_i \left[\mathbf{u}\cdot \nabla \rho+\frac{\mathbf{c}_i \cdot (\mathbf{F}+\mathbf{F}_c)}{c_s^2}+\frac{\mathbf{Q}:\left(\mathbf{c}_i \mathbf{c}_i-c_s^2 \mathbf{I}\right)}{2 c_s^4}\right],
	\end{equation}
\end{subequations}
where $\mathbf{F}_c=-\nabla \cdot(\mathbf{m}^C\mathbf{u}-\mathbf{u}\mathbf{m}^C)$/2 and $\mathbf{Q}=(\mathbf{F}+\mathbf{F}_c)\mathbf{u}+\mathbf{u}(\mathbf{F}+\mathbf{F}_c)+\partial_t(\mathbf{m}^C\mathbf{u}+\mathbf{u}\mathbf{m}^C)/2+c_s^2(\mathbf{u}\nabla\rho+\nabla\rho\mathbf{u})$.

In addition, the order parameter $C$, fluid velocity $\mathbf{u}$, the hydrodynamic pressure $p$ and electrical potential $\phi$ are calculated by \cite{Zhan2022,Chai2008}
\begin{subequations}
	\begin{equation}
		C=\sum_{i} f_i,
	\end{equation}
	\begin{equation}
		\rho \mathbf{u}=\sum_{i}\mathbf{c}_i g_i +\frac{\Delta t}{2}(\mathbf{F}+\mathbf{F}_c),
	\end{equation}
	\begin{equation}
		p=\frac{c_s^2}{1-\omega_0}\left\{\sum_{i\not =0} g_i + \frac{1}{2}\Delta t\mathbf{u}\cdot\nabla\rho + s_0(\rho,C,\mathbf{u}) + \frac{1}{3c^2}\Delta t[2(\mathbf{F}+\mathbf{F}_c)\cdot\mathbf{u}+\partial_t(\mathbf{m}^C \cdot \mathbf{u})]\right\},
	\end{equation}
	\begin{equation}
		\phi=\sum_{i \not= 0}\frac{1}{1-\omega_0}h_i.
	\end{equation}
\end{subequations}

It should be noted that, to calculate the gradient and Laplacian terms in the LB models, the following second-order isotropic central schemes are used,
\begin{subequations}
	\begin{equation}
		\nabla \xi(\mathbf{x})=\sum_{i \not= 0}\frac{\omega_i \mathbf{c}_i \xi(\mathbf{x}+\mathbf{c}_i \Delta t)}{c_s^2 \Delta t},
	\end{equation}
	\begin{equation}
		\nabla^2 \xi(\mathbf{x})=\sum_{i \not= 0}\frac{2\omega_i [\xi(\mathbf{x}+\mathbf{c}_i \Delta t)-\xi(\mathbf{x})]}{c_s^2 \Delta t^2}.
	\end{equation}
\end{subequations}

\subsubsection{Discrete velocity model}\label{DdQq}
In this study, we only consider two-dimensional problems, and the nine-velocity (D2Q9) lattice model \cite{Qian1992} is adopted for both phase and flow fields, while for simplicity, the five-velocity (D2Q5) model \cite{Liu2019} is applied for the electric field. In the D2Q9 model, the discrete velocity $\mathbf{c}_i$, weight coefficient $\omega_i$, orthogonal transformation matrix $\mathbf{M}_f=\mathbf{M}_g$, and diagonal relaxation matrices $\mathbf{S}_f$ and $\mathbf{S}_g$ are given by
\begin{subequations}
	\begin{equation}
		\begin{split}
			\mathbf{c}_i=\left(\begin{array}{ccccccccc}   
				0 & 1 & 0 & -1 &  0 & 1 & -1 & -1 &  1\\  
				0 & 0 & 1 &  0 & -1 & 1 &  1 & -1 & -1\\  
			\end{array}\right)c,
		\end{split}
	\end{equation}
	\begin{equation}
		\begin{split}
			\omega_i=\left(\begin{array}{ccccccccc}   
				\dfrac{4}{9} & \dfrac{1}{9} & \dfrac{1}{9} & \dfrac{1}{9} &  \dfrac{1}{9} & \dfrac{1}{36} & \dfrac{1}{36} & \dfrac{1}{36} &  \dfrac{1}{36}\\  
			\end{array}\right),
		\end{split}
		\label{wi}
	\end{equation}
	\begin{equation}      
		\begin{split}
			\mathbf{M}_f=\mathbf{M}_g=\left( \begin{array}{rrrrrrrrrrrrrrrrr}   
				1 &&  1  &&  1  &&  1  &&  1  && 1  &&  1  &&  1  &&   1\\ 
				-4 && -1  && -1  && -1  && -1  && 2  &&  2  &&  2  &&   2\\ 
				4 && -2  && -2  && -2  && -2  && 1  &&  1  &&  1  &&   1\\  
				0 &&  1  &&  0  && -1  &&  0  && 1  && -1  && -1  &&   1\\ 
				0 && -2  &&  0  &&  2  &&  0  && 1  && -1  && -1  &&   1\\  
				0 &&  0  &&  1  &&  0  && -1  && 1  &&  1  && -1  &&  -1\\  
				0 &&  0  && -2  &&  0  &&  2  && 1  &&  1  && -1  &&  -1\\  
				0 &&  1  && -1  &&  1  && -1  && 0  &&  0  &&  0  &&   0\\  
				0 &&  0  &&  0  &&  0  &&  0  && 1  && -1  &&  1  &&  -1\\  
			\end{array}\right), 
		\end{split}
	\end{equation}
	\begin{equation}
		\mathbf{S}_f=\mathbf{diag}\left(s_0^f,s_1^f,s_2^f,s_3^f,s_4^f,s_5^f,s_6^f,s_7^f,s_8^f\right),
	\end{equation}
	\begin{equation}
		\mathbf{S}_g=\mathbf{diag}\left(s_0^g,s_1^g,s_2^g,s_3^g,s_4^g,s_5^g,s_6^g,s_7^g,s_8^g\right),
	\end{equation}
\end{subequations}
where $c=\Delta x/\Delta t$ is the lattice speed with $\Delta x$ and $\Delta t$ being the lattice spacing and time step, the relaxation parameters $s_3^f$ and $s_5^f$ are related to the mobility, and are given by $s_3^f=s_5^f=1/(M/c_s^2\Delta t+0.5)$. Similarly, $s_7^g$ and $s_8^g$ are related to the kinematic viscosity $v=\mu/\rho$, and can be determined by $s_7^g=s_8^g=1/(v/c_s^2\Delta t+0.5)$, $c_s=c/\sqrt{3}$ is the sound speed. Additionally, the other relaxation parameters are set as $1$ if not specified. In D2Q5 model, the discrete velocity $\mathbf{c}_i$ and weight coefficient $\omega_i$, orthogonal transformation matrix $\mathbf{M}_h$, and diagonal relaxation matrix $\mathbf{S}_h$ can be given as
\begin{subequations}
	\begin{equation}
		\begin{split}
			\mathbf{c}_i=\left(\begin{array}{ccccccccc}   
				0 & 1 & 0 & -1 &  0\\  
				0 & 0 & 1 &  0 & -1\\  
			\end{array}\right)c,
		\end{split}
	\end{equation}
	\begin{equation}
		\begin{split}
			\omega_i=\left(\begin{array}{ccccccccc}   
				\dfrac{1}{3} & \dfrac{1}{6} & \dfrac{1}{6} & \dfrac{1}{6} &  \dfrac{1}{6} \\  
			\end{array}\right),
		\end{split}
	\end{equation}
	\begin{equation}      
		\begin{split}
			\mathbf{M}_h=\left( \begin{array}{rrrrrrrrrrrrrrrrr}   
				1 &&  1  &&  1  &&  1  &&  1\\ 
				0 && 1  && 0  && -1  && 0\\ 
				0 && 0  && 1  && 0  && -1\\  
				0 &&  1  &&  -1  && 1  &&  -1\\ 
				-4 && 1  &&  1  &&  1  &&  1\\    
			\end{array}\right), 
		\end{split}
	\end{equation}
	\begin{equation}
		\mathbf{S}_h=\mathbf{diag}\left(s_0^h,s_1^h,s_2^h,s_3^h,s_4^h\right),
	\end{equation}
\end{subequations}
where the relaxation parameters $s_1^h$ and $s_2^h$ are related to the conductivity through the relation $s_1^h=s_2^h=1/(\sigma/c_s^2\Delta t+0.5)$, and other relaxation parameters are fixed to be 1.

\section{Numerical validations}\label{validation}
In this section, we validate the LB method by two benchmark problems, and the half-way bounce-back scheme \cite{Ladd1994,Ladd1994_2} is adopted to treat the no-slip and no-flux boundary conditions, while the general bounce-back scheme \cite{Zhang2012} is adopted to treat the Dirichlet boundary condition. 
\subsection{The droplet spreading on an ideal wall}\label{validation1}
We first consider the spreading of a droplet on an ideal wall to test the capability of the LB method in predicting the contact angle. For this problem, the density ratio ($\rho_l/\rho_g$) and dynamic viscosity ratio ($\mu_l/\mu_g$) are fixed to be 1000 and 100, the computational domain is $[0,200]\times[0,100]$, in which a semicircular droplet with the radius $R_0=35$ is deposited on the bottom wall. The other parameters in numerical simulations are set as $\Delta x=\Delta t=1,\ W = 5,\ \rho_g=1,\ v_g=0.1,\ \gamma=0.2$ and $M=0.1$, the initial distribution of the order parameter is given by
\begin{equation}
	C(x,y)=\frac{C_l+C_g}{2}+\frac{C_l-C_g}{2}\mathrm{tanh}\frac{2 \big(R_0-\sqrt{(x-100)^2+y^2} \big)}{W}.
\end{equation}
In addition, for the phase field, the periodic boundary condition is applied in the horizontal direction, while the wetting and no-flux boundary conditions are imposed at the bottom and top boundaries. For the flow field, the periodic boundary condition is used in the horizontal direction, while the no-slip boundary condition is adopted at the bottom and top boundaries.
\begin{figure}[h]
	\centering
	\subfigure[]{
		\begin{minipage}[b]{0.45\linewidth}
			\centering
			\hspace{-5mm}
			\includegraphics[width=2.0in]{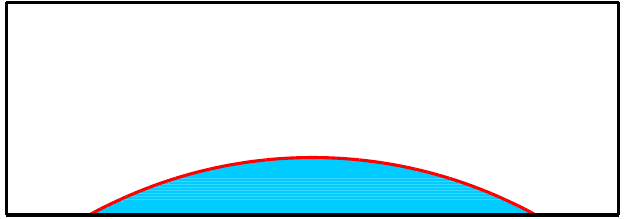}
	\end{minipage}}
	\subfigure[]{
		\begin{minipage}[b]{0.45\linewidth}
			\centering
			\hspace{-5mm}
			\includegraphics[width=2.0in]{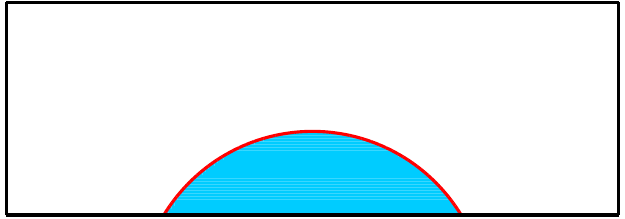}
	\end{minipage}}
	
	\subfigure[]{
		\begin{minipage}[b]{0.45\linewidth}
			\centering
			\hspace{-4mm}
			\includegraphics[width=2.0in]{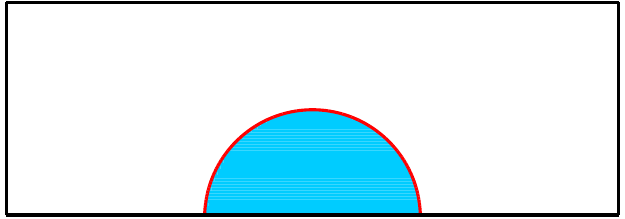}
	\end{minipage}}
	\subfigure[]{
		\begin{minipage}[b]{0.45\linewidth}
			\centering
			\hspace{-4mm}
			\includegraphics[width=2.0in]{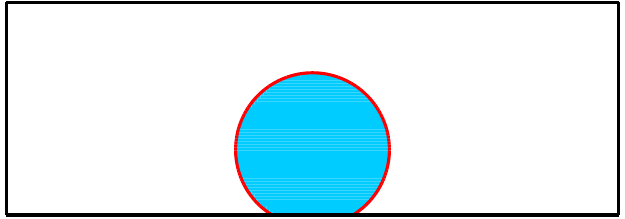}
	\end{minipage}}
	\caption{The predicted equilibrium shapes of the droplet under different contact angles [(a) $\theta=30^{\circ}$, (b) $\theta=60^{\circ}$, (c) $\theta=90^{\circ}$, (d) $\theta=150^{\circ}$].}	
	\label{picture-ContactAngle}
\end{figure}

\begin{table}[h]
	\centering
	\caption{A comparison of the theoretical and numerical results of the contact angle.}
	\vspace{2pt}
	\begin{tabular}{ccccccccccccc}\hline \hline
		Theoretical     && $30^{\circ}$ && $45^{\circ}$ && $60^{\circ}$ && $90^{\circ}$ && $120^{\circ}$ && $150^{\circ}$ \\
		\hline
		Numerical && $29.3^{\circ}$ && $44.2^{\circ}$ && $59.3^{\circ}$ && $89.5^{\circ}$ && $119.9^{\circ}$ && $150.0^{\circ}$ 
		\\ \hline \hline
	\end{tabular}
	\label{table-ContactAngle}
\end{table}
We perform some simulations, and present the results in Fig. \ref{picture-ContactAngle} which shows the equilibrium shapes of the droplets under different contact angles. From this figure, one can see that the droplet on the solid wall can form different steady-state patterns, which depends on the specified contact angle. Based on the equilibrium state of the droplet, we can also numerically calculate the contact angle by using the formula  $\theta=2\mathrm{arctan}(2H/L)$, where $L$ and $H$ are the spreading length and height of the droplet on the solid surface, respectively. As seen from Table \ref{table-ContactAngle}, the predicted values of contact angles are in good agreement with theoretical ones.
\begin{figure}[h]
	\centering
	\includegraphics[width=2.5in]{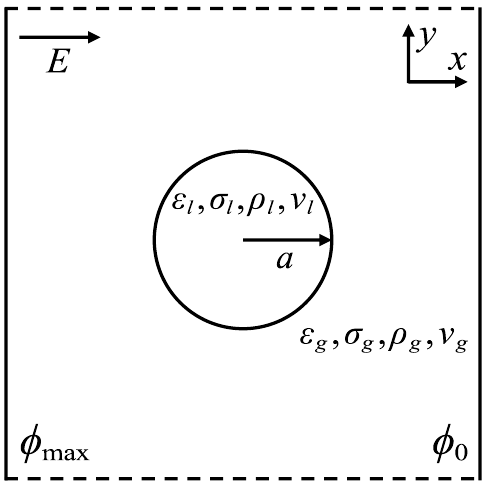}
	\caption{Schematic of a droplet suspending in another immiscible liquid under an external electric field.}
	\label{picture-gkt_bx}
\end{figure}

\subsection{The deformation of a leaky dielectric droplet in an external electric field}\label{validation2}
The deformation of a leaky dielectric droplet subjected to an external electric field is also used to test the capability of the LB method for the problems with the multi-physical fields. As shown in Fig. \ref{picture-gkt_bx} where a single droplet (marked as $l$) with the radius $a$ is suspended in another immiscible fluid (marked as $g$), when an electric field is applied, the droplet deforms into either a prolate or oblate shape with the internal and external circulation flows, which are induced by the electrical stress ($\mathbf{F}_E$) acting on the interface. To quantify the droplet deformation, we adopt the following deformation factor, 
\begin{equation}
	D=\frac{\hat{L}-\hat{H}}{\hat{L}+\hat{H}},
\end{equation}
where $\hat{L}$ and $\hat{H}$ are the end-to-end lenghts of the stable droplet measured along and perpendicularly to the direction of the electric field, respectively. Based on the classic leaky dielectric theory \cite{Taylor1966}, an approximate expression of deformation factor can be given by
\begin{equation}
	D=\frac{9 Ca_E}{16}\frac{f_T(R,S,B)}{(2+R)^2},
	\label{Taylor}
\end{equation}
where $f_T(R,S,B)=R^2+1-2S+\frac{3}{5}(R-S)\frac{2+3B}{1+B}$, which is a function of the conductivity ratio ($R=\sigma_l/\sigma_g$), the permittivity ratio ($S=\varepsilon_l/\varepsilon_g$) and the dynamic viscosity ratio ($B=\mu_l/\mu_g$). The electric capillary number ($Ca_E=\varepsilon_g E_0^2 a / \gamma$) is used to reflect the influence of the electric field. Similarly, Feng \cite{Feng2002} also proposed a first-order approximate formula for 2D droplet deformation under a uniform electric field,
\begin{equation}
	D=\frac{Ca_E f_F(R,S)}{3(1+R)^2},
	\label{Feng}
\end{equation}
where $f_F(R,S)=R^2+R+1-3S$. Moreover, when $D>0$, the droplet is prolate, while if $D<0$, the droplet is oblate.
\begin{figure}[h]
	\centering
	\subfigure[]{
		\begin{minipage}[b]{0.45\linewidth}
			\centering
			\hspace{-3mm}
			\includegraphics[width=2.5in]{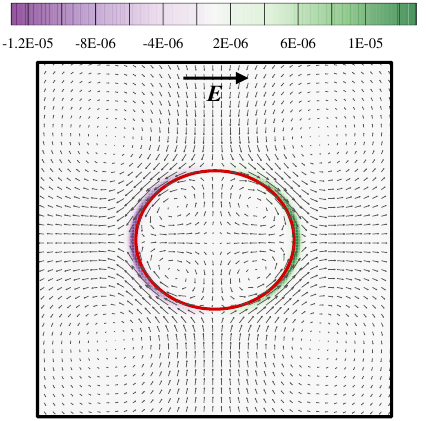}
		\end{minipage}
		\label{picture-bx_0.5}}
	\subfigure[]{
		\begin{minipage}[b]{0.45\linewidth}
			\centering
			\hspace{-3mm}
			\includegraphics[width=2.53in]{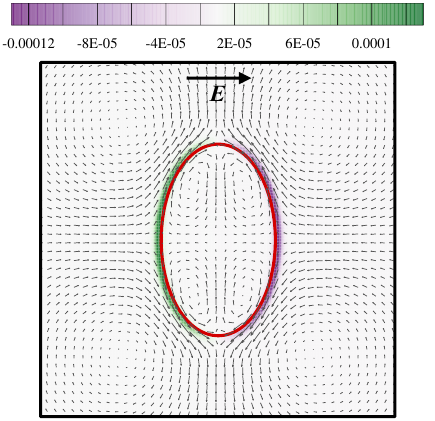}
		\end{minipage}
		\label{picture-bx_60}}
	\caption{The deformation of the leaky dielectric droplet and vortex direction [(a) $(R,S)=(5,0.5)$, (b) $(R,S)=(5,60)$].}
	\label{picture-bx}
\end{figure}

\begin{table}[h]
	\centering
	\caption{A comparison of the numerical and theoretical results on the deformation of the leaky dielectric droplet.}
	\renewcommand{\arraystretch}{1.0} \tabcolsep 10pt
	\vspace{2pt}
	\begin{tabular}{cccccccccc}\hline \hline
		& & & \multicolumn{4}{c}{Deformation($D$)} \\ \cline{4-7}
		$R$ & $S$ & $Ca_E$ & Eq.\ (\ref{Taylor})  & Eq.\ (\ref{Feng}) & Liu \cite{Liu2019} & Present \\ \hline
		5 & 5 & 0.2 & 0.03670 & 0.02960 & 0.03524 & 0.03535 \\
		5 & 60 & 0.2 & -0.40520 & -0.27590 & -0.25708 & -0.25296 \\
		1 & 2 & 0.2 & -0.04380 & -0.05000 & -0.04751 & -0.04997 \\
		50 & 2 & 0.2 & 0.10690 & 0.06520 & 0.10756 & 0.10689 \\
		1.75 & 3.5 & 0.1 & -0.02230 & -0.02070 & -0.02232 & -0.02119 \\
		3.25 & 3.5 & 0.1 & 0.00850 & 0.00800 & 0.00833 & 0.00874 \\
		4.75 & 3.5 & 0.1 & 0.02280 & 0.01800 & 0.01953 & 0.02092 \\ \hline \hline
	\end{tabular}
	\label{Table-D}
\end{table}

In our simulations, the computational domain is $L_x \times L_y=400\times400$ with $\Delta x=\Delta t=1$, other parameters are set to be $a=50,\ W = 5\Delta x,\ \rho_l=\rho_g=1,\ v_l=v_g=0.1,\ \gamma=0.001,\ M=0.1,\ \sigma_g=0.1,\ \varepsilon_g=0.0004,\ E_0=\sqrt{Ca_E \gamma/(a\varepsilon_g)},\ \phi_0=0$ and $\phi_{max}=\phi_0+E_0L_x$. The initial distribution of the order parameter is given by
\begin{equation}
	C(x,y)=\frac{C_l+C_g}{2}+\frac{C_l-C_g}{2}\mathrm{tanh}\frac{2\left[a-\sqrt{(x-L_x/2)^2+(y-L_y/2)^2}\right]}{W}.
\end{equation}
The boundary conditions are considered as follows. For the phase and flow fields, the no-flux and no-slip boundary conditions are imposed on the left and right walls, while for the electric potential, the Dirichlet boundary conditions with $\phi_{max}$ and $\phi_0$ are applied on the left and right walls, respectively. In addition, the periodic boundary conditions are employed on the top and bottom boundaries.

As seen from Figure \ref{picture-bx} where $Ca_E=0.2$, the droplet can form prolate [see Fig. \ref{picture-bx_0.5}] and oblate [see Fig. \ref{picture-bx_60}] patterns under the conditions of  $(R,S)=(5,0.5)$ and $(R,S)=(5,60)$. In addition, it is also found that there are four symmetric vortices inside and outside the droplet, and when $R>S$, the flow direction is from the equator to the poles along the droplet surface, while if $R<S$, the flow direction is from the poles to the equator along the droplet surface. We also present the distribution of the volume charge density $\rho_e$ in Fig. \ref{picture-bx}, from which one can see that when $R>S$, the left half of the droplet has a negative charge and the right half has a positive charge, while the distribution is opposite if $R<S$. Furthermore, we present a comparison of deformation factor among the present results, theoretical solutions [Eqs. (\ref{Taylor}) and (\ref{Feng})], and numerical data \cite{Liu2019} in Table \ref{Table-D}. This table demonstrates that our numerical results are not only in good agreement with the theoretical solutions for the droplet with small deformation, but also close to the previous numerical data \cite{Liu2019}. Here it should be noted that for the droplet with large deformation, the theoretical solutions are not accurate, and there are some differences between the numerical and theoretical data, as reported in the previous work \cite{Liu2019}.

\section{Numerical results and discussion}\label{Results}
In this section, we investigate the electrohydrodynamics of a droplet pair on the solid substrate by the LB method. The schematic of the problem is shown in Fig. \ref{picture-gkt_xm} where a pair of semicircular leaky dielectric droplets (marked as $l$) with the same radius $a$ is initially placed on the solid substrate, and is also surrounded by another immiscible fluid (marked as $g$). The distance between the centers of the two droplets is $4a$, and an external electric field is applied in the horizontal direction, In the following simulations, the computational domain is $L_x \times L_y=640\times160$, the parameters are given by $a=40,\ W = 5,\ \rho_l=\rho_g=1,\ v_l=v_g=0.05,\ \gamma=0.004,\ M=0.1,\ R=\sigma_l/\sigma_g=5,\ \sigma_g=1/6,\ Ca_E=0.5,\ \varepsilon_g=0.005,\ E_0=\sqrt{Ca_E \gamma/(a\varepsilon_g)}=0.1,\ \phi_0=0$ and $\phi_{max}=\phi_0+E_0L_x$. The initial order parameter is set as
\begin{equation}
	C(x,y)=
	\begin{cases}
		\frac{C_l+C_g}{2}+\frac{C_l-C_g}{2}\mathrm{tanh}\frac{2\left[a-\sqrt{[x-(L_x/2-2a)]^2+y^2}\right]}{W}, & 0 \leq x \leq L_x/2,0 \leq y \leq L_y,\\
		\frac{C_l+C_g}{2}+\frac{C_l-C_g}{2}\mathrm{tanh}\frac{2\left[a-\sqrt{[x-(L_x/2+2a)]^2+y^2}\right]}{W}, & L_x/2 \leq x \leq L_x,0 \leq y \leq L_y.
	\end{cases}
\end{equation}
In addition, for the phase field, the no-flux boundary is applied to the left, right and top boundaries, while the wetting boundary condition is imposed on the bottom boundary. For the flow field, the no-slip boundary condition is adopted for all four boundaries. For the electric field, the Dirichlet boundary conditions with $\phi_{max}$ and $\phi_0$ are used at the left and right boundaries, while the no-flux boundary conditions are imposed on the bottom and top boundaries.

\subsection{A grid-independence test}\label{grid}
Before conducting any discussion, a grid-independence test is first performed. As illustrated in Fig. \ref{picture-gkt_xm}, the inner-end distance of the two droplets is $L_1=2a$, and its dimensionless form is given as $L_1^*=L_1/2a$. Additionally, we also define a dimensionless time $t^*=n\Delta t/t_c$, where $n$ denotes the number of time steps, $t_c=\mu_g/(\varepsilon_g E_0^2)$ represents the characteristic time, with  $t_c=1000$ taken as a reference value. We consider the case with $\theta=90^{\circ}$ and $S=\varepsilon_l/\varepsilon_g=1.5$, and find that the two droplets initially attract each other, and subsequently remain stable. Table \ref{table-dlxyz} presents the stable values of $L_1^*$ under different grid resolutions, and the relative errors are calculated based on the reference value with a grid of $N_x \times N_y=1600\times400$. As shown in the table, the grid resolution $1280\times320$ can give grid-independent results, and the relative error is less than $1\%$. This grid resolution is adopted in the following simulations considering computational accuracy and efficiency.
\begin{table}[h]
	\centering
	\caption{The relative errors of $L_1^*$ under different grid resolutions.}
	\vspace{2pt}
	\begin{tabular}{ccccccccccccc}\hline \hline
		Grid size     && $640\times160$ && $1280\times320$ && $1600\times400$ \\ \hline
		$L_1^*$     && 0.157 && 0.164 && 0.165 \\ \hline
		Relative error (\%) && 4.85 && 0.61 && -- \\ \hline \hline
	\end{tabular}
	\label{table-dlxyz}
\end{table}

\begin{figure}[t]
	\centering
	\includegraphics[width=4.0in]{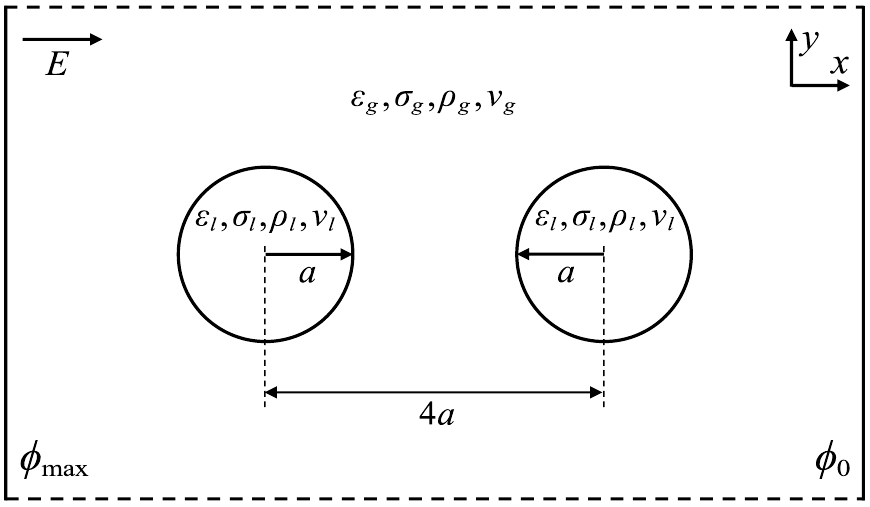}
	\caption{Schematic of a droplet pair suspending in another immiscible liquid under an external electric field. The subscripts $l$ and $g$ represent the inner and outer fluids of droplets, respectively.}
	\label{picture-gkt_zj}
\end{figure}

\begin{figure}[t]
	\centering
	\subfigure[]{
		\begin{minipage}[b]{0.46\linewidth}
			\centering
			\hspace{-7.2mm}
			\includegraphics[width=3.0in]{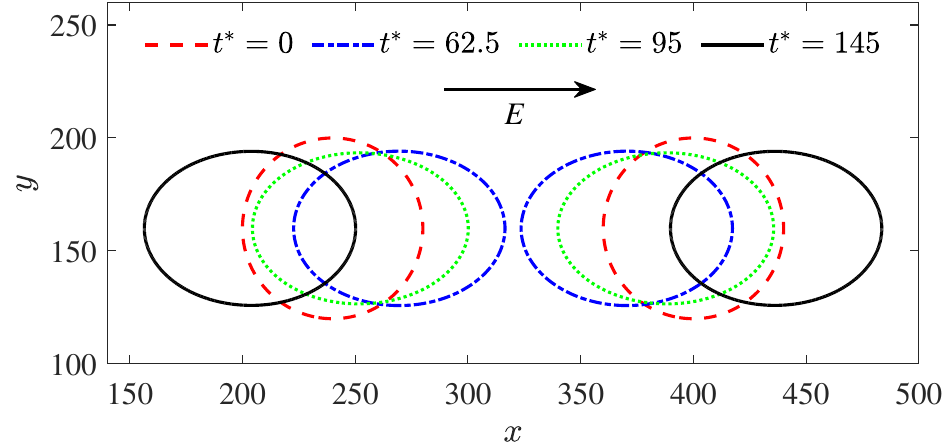}
		\end{minipage}
		\label{picture-bj_suspension}}
	\subfigure[]{
		\begin{minipage}[b]{0.46\linewidth}
			\centering
			\hspace{-8.4mm}
			\includegraphics[width=3.0in]{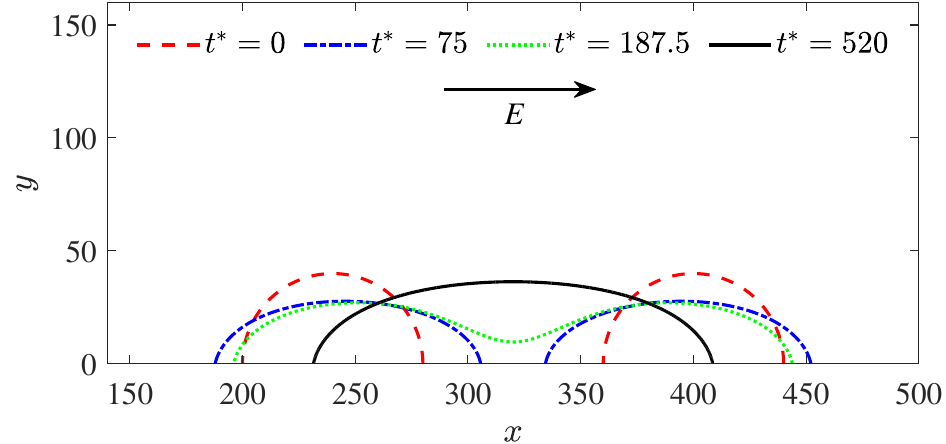}
	\end{minipage}
      	\label{picture-bj2}}

	\caption{The electrohydrodynamics of the droplet pair in two different situations [(a) Suspension of the droplet pair, (b) the droplet pair on the solid substrate].}
	\label{picture-bj}
\end{figure}

To investigate the influence of wettability on the electrohydrodynamics of a droplet pair under an applied electric field, two different cases are considered, the first one is a droplet pair suspended in another fluid (see Fig. \ref{picture-gkt_zj}), and the second one is a droplet pair placed on a solid substrate (see Fig. \ref{picture-gkt_xm}). For the first case, the periodic boundary conditions are applied to the top and bottom boundaries, and the other boundary conditions and parameters are the same as those for the second case. From Fig. \ref{picture-bj_suspension}, it can be observed that when two droplets are suspended, they initially approach each other, and subsequently move apart. This is because during this process, the droplets are subjected to both electrostatic force and electrically driven hydrodynamic force \cite{Pohl1978,Baygents1998,Dong2018}. To explain this phenomenon more clearly, we first introduce two time scales for tangential electric stress $\tau_\mu \sim d^2 \rho / \mu$ and the Rayleigh inertial capillary $\tau_\rho^\gamma=\sqrt{\rho a^3 / \gamma}$, then the timescale for the electric-stress-driven flow near one droplet to influence the other can be defined as $\tau_\mu /  \tau_\rho^\gamma \sim d^2 / (a^2 Oh)$, where the Ohnesorge number $Oh$ is defined by $Oh=\mu/\sqrt{\rho a\gamma}$, $d$ is the centre-to-centre distance between the droplets \cite{Das2021}. For the case with a small $Oh=0.125$, the droplet pair initially responds to the electrostatic interaction, which can be modeled as the force between two dipoles located at the droplet centers. The horizontal ($F_h$) and vertical ($F_v$) components of this force take the following forms \cite{Chiesa2005},
\begin{equation}
	F_h=\frac{12 \pi \varepsilon_g E_0^2 a^6 \hat{B}^2}{d^4}[3(\mathrm{cos}\alpha)^2-1],\ \ \ \ F_v=-\frac{12 \pi \varepsilon_g E_0^2 a^6 \hat{B}^2}{d^4} \mathrm{sin}2\alpha,
	\label{dxhzy}
\end{equation}
where $\alpha$ is the angle between the direction of the electric field and the line along the centers of two droplets, and $\hat{B}=1-3/(S+2)$. When the electric field is applied in the horizontal direction ($\alpha=0^{\circ}$), the vertical component of electric force is $0$, and the horizontal component has a positive sign. This explains why the droplets initially approach each other. Additionally, due to the fact that the tangential component of the electrical stress is non-zero at the interface, some vortices are generated inside and outside the droplet, and the tangential velocity at the interface can be expressed as \cite{Baygents1998}
\begin{equation}
	u_{\hat{\theta}}=-\frac{9(R-S)}{5(1+B)(R+2)^2}\mathrm{cos} \hat{\theta} \mathrm{sin} \hat{\theta},
	\label{sdlxhzy}
\end{equation}
where $\hat{\theta}$ is the angle between the direction of applied electric field and the line connecting the droplet center to a point on the interface, and from Eq. (\ref{sdlxhzy}), one can see that there are two types circulation, with their directions are determined by the value of $\hat{\theta}$ and sign of the term $R-S$. As pointed out in Ref. \cite{Baygents1998}, when $\alpha=0^{\circ}$, the vortical circulations around one droplet push the other droplet away if $R>S$, and the circulations around one droplet pull the other one closer if $R<S$. With the evolution of the flow field, the dominant electrically driven hydrodynamic interaction causes the droplets to gradually separate from each other.

However, for the case of a droplet pair placed on the substrate, it initially attracts to each other, and subsequently coalesces [see Fig. \ref{picture-bj2}]. This indicates that the wettability has an important influence on the electrohydrodynamics and final state of the droplet pair, which will be considered in the following part.
\begin{figure}[h]
	\centering
	\subfigure[Phase diagram of a droplet pair on the solid substrate]{
		\begin{minipage}[b]{0.56\linewidth}
			\centering
			\hspace{-6.1mm}
			\includegraphics[width=3.8in]{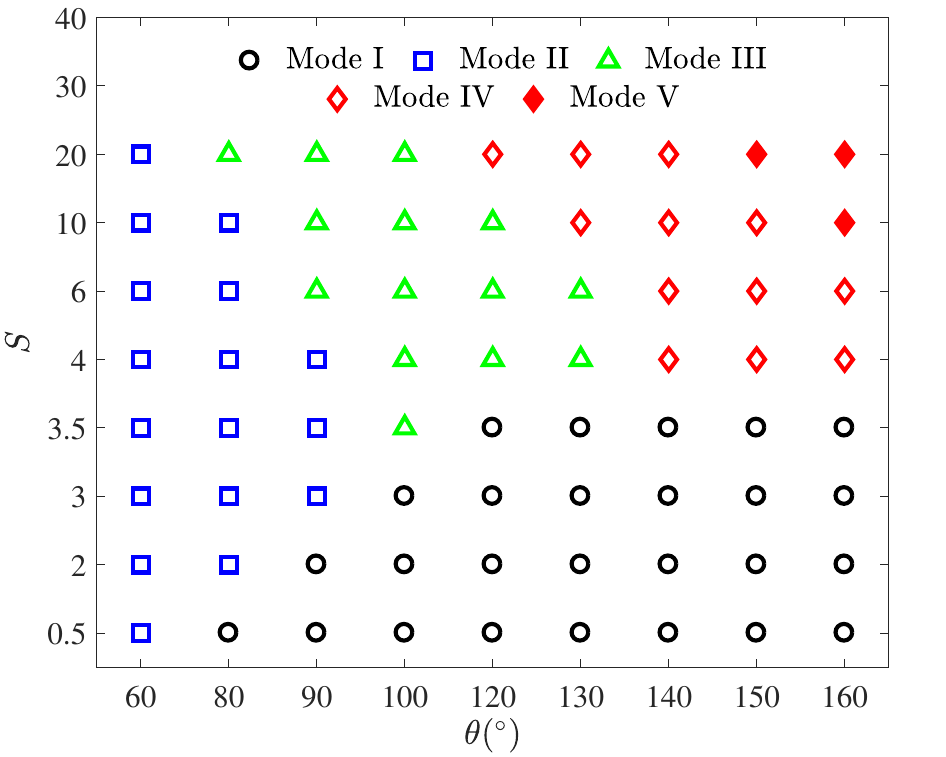}
		\end{minipage}
		\label{picture-xt1}}
	\subfigure[Modes IV (top) and V (bottom)]{
		\begin{minipage}[b]{0.35\linewidth}
			\centering
			\hspace{-5.2mm}
			\vspace{1mm}
			\includegraphics[width=2.1in]{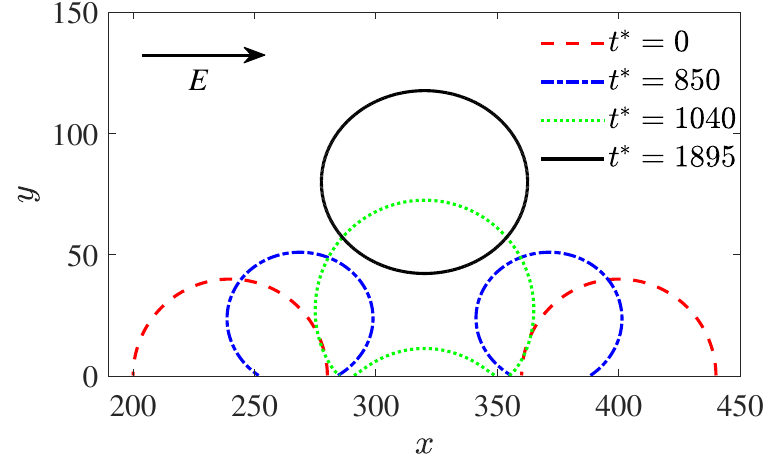}\vspace{20pt}
			
			\hspace{-4.2mm}
			\includegraphics[width=2.1in]{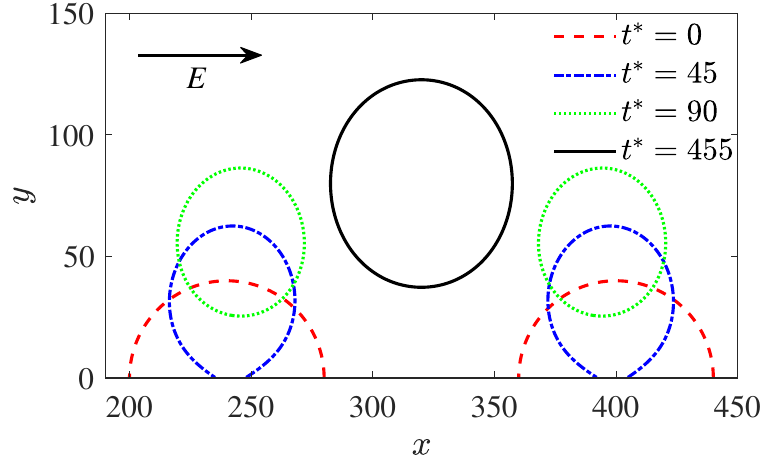}
			\vspace{3mm}
		\end{minipage}
		\label{picture-mode4_5}}
	
	\subfigure[Modes I (left), II (middle) and III (right)]{
		\begin{minipage}[b]{1.0\linewidth}
			\centering
			\hspace{-6.2mm}
			\includegraphics[width=2.1in]{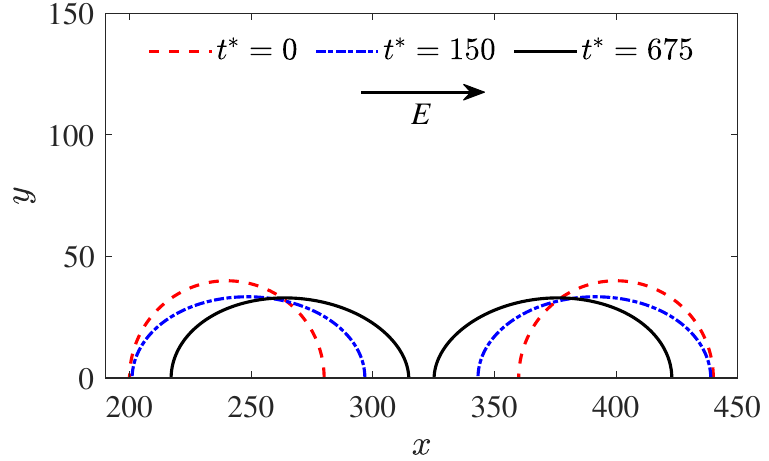}
			\includegraphics[width=2.1in]{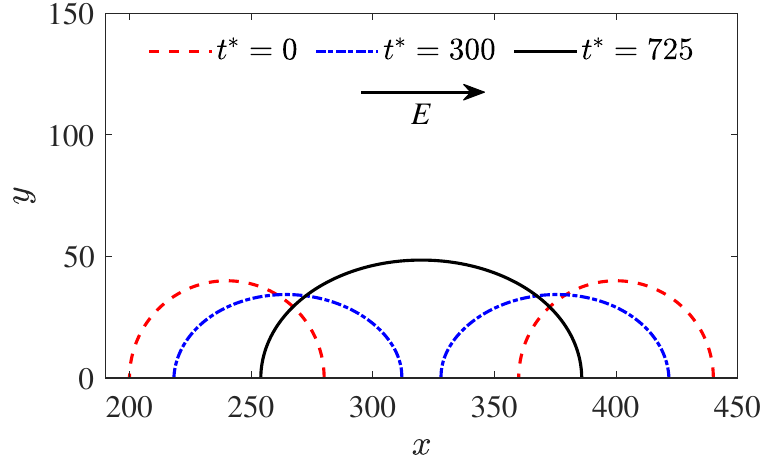}
			\includegraphics[width=2.1in]{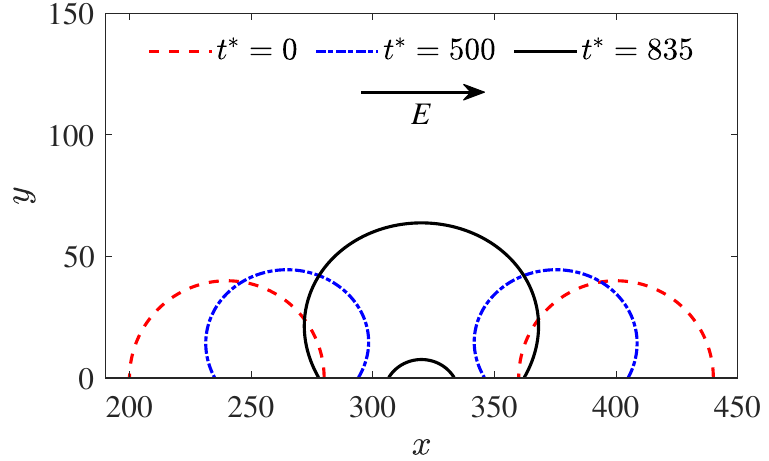}
		\end{minipage}
		\label{picture-mode1_2_3}}
	\caption{The modes of the droplet pair at different contact angles ($\theta$) and permittivity ratios ($S$) [{\color{black}{$\circ$}}: attraction without coalescence, {\color{blue}{$\square$}}: attraction with coalescence, {\color{green}{$\triangle$}}: coalescence with bubble entrapment, {\color{red}{$\lozenge$}}: coalescence followed by suspension, {\color{red}{$\blacklozenge$}}: suspension followed by coalescence].}
\end{figure}
\begin{figure}[]
	\centering
	\subfigure[]
	{
		\begin{minipage}[b]{0.31\linewidth}
			\hspace{-2mm}
			\includegraphics[width=2.0in]{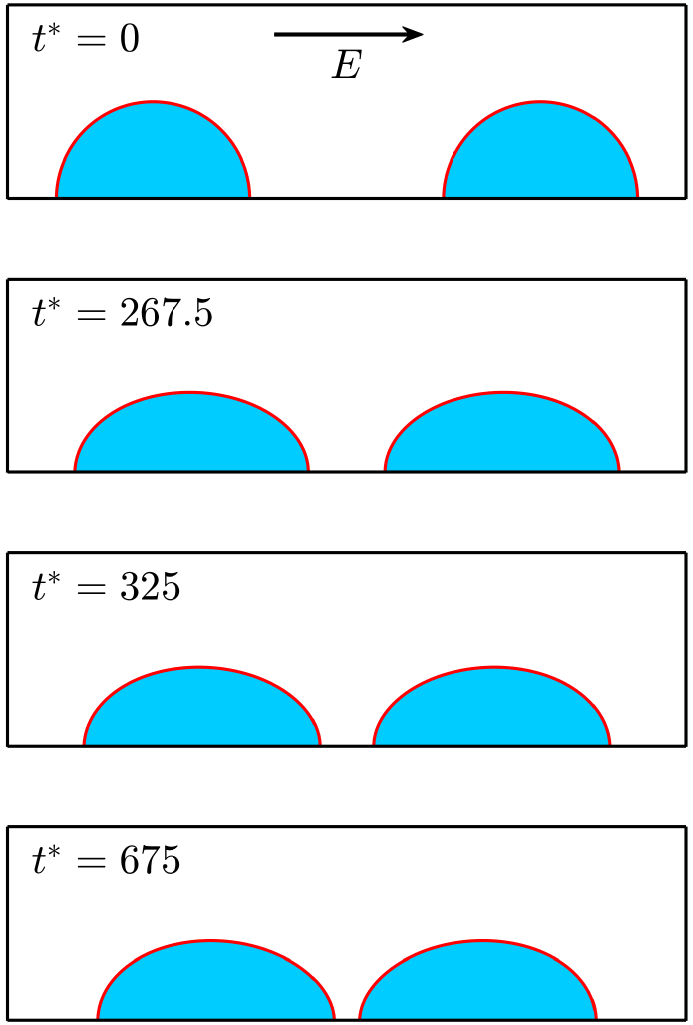}
		\end{minipage}
		\label{picture-90_2}
	}
	\subfigure[]
	{
		\begin{minipage}[b]{0.31\linewidth}
			\hspace{-2mm}
			\includegraphics[width=2.0in]{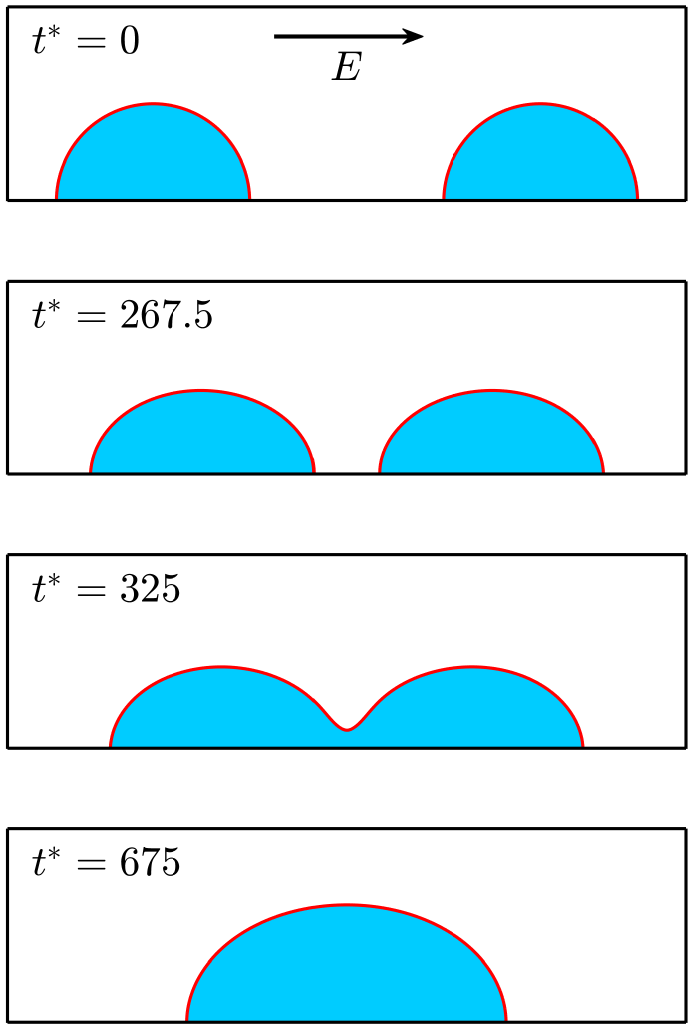}
		\end{minipage}
		\label{picture-90_4}
	}
	\subfigure[]
	{
		\begin{minipage}[b]{0.31\linewidth}
			\hspace{-2mm}
			\includegraphics[width=2.0in]{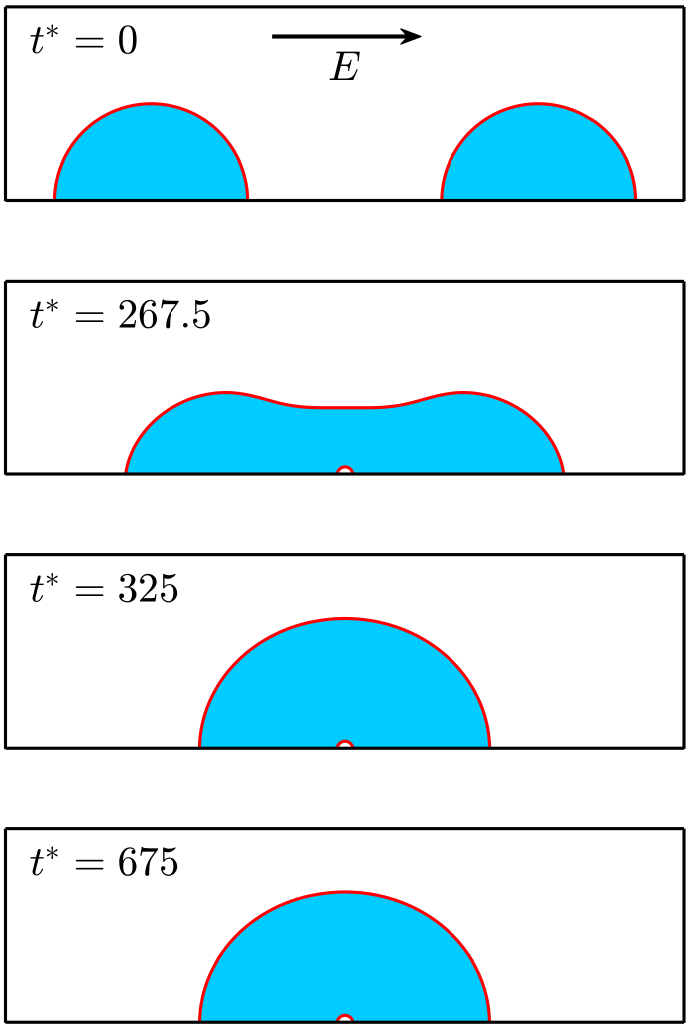}
		\end{minipage}
		\label{picture-90_8}
	}
	\caption{The snapshots of the droplet pair at different dimensionless times ($t^*$) [(a) $S=2$, (b) $S=4$, (c) $S=8$].}
	\label{picture-sp_90}
\end{figure}
\begin{figure}[]
	\centering
	\subfigure[]{
		\begin{minipage}[b]{0.475\linewidth}
			\centering
			\hspace{-3mm}
			\includegraphics[width=3.0in]{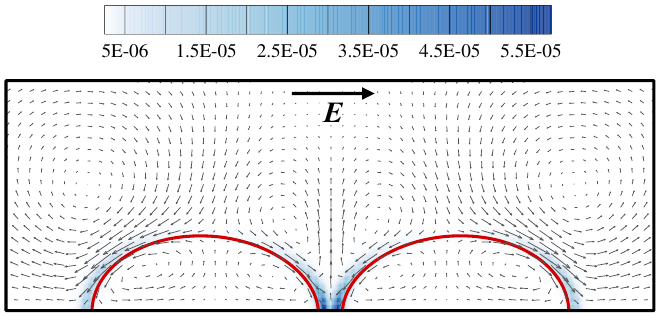}
		\end{minipage}
		\label{picture-sp_lxt_1}}
	\subfigure[]{
		\begin{minipage}[b]{0.475\linewidth}
			\centering
			\hspace{-3mm}
			\includegraphics[width=3.0in]{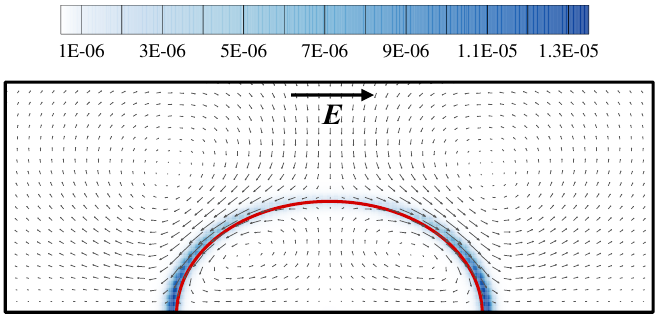}
		\end{minipage}
		\label{picture-sp_lxt_2}}
	
	\subfigure[]{
		\begin{minipage}[b]{0.475\linewidth}
			\centering
			\hspace{-3mm}
			\includegraphics[width=3.0in]{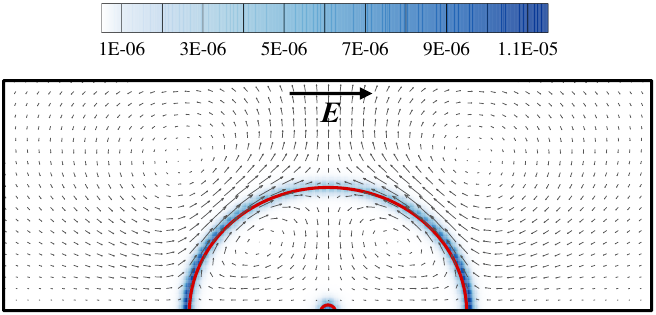}
		\end{minipage}
		\label{picture-sp_lxt_3}}
	\caption{The distribution of electrical stress $\mathbf{F}_E$ at different permittivity ratios [(a) $S=2$, (b) $ S=4$, (c) $S=8$].}
	\label{picture-sp_lxt} 
\end{figure}

\subsection{The droplet pair in a horizontal external electric field}
Based on the above analysis, the permittivity ratio ($S$) significantly affects the droplet deformation, as well as the intensity of electrostatic interaction and hydrodynamic interaction. Besides, the wettability also exerts an apparent impact on the interaction between two droplets. In the following, we will investigate the effects of permittivity ratio and wettability on the interaction of droplet pair placed on a solid substrate under a horizontal electric field (see Fig. \ref{picture-gkt_xm}). We first present the phase diagram for the modes of the droplet pair in Fig. \ref{picture-xt1}. As seen from this figure, there are five different modes under different values of permittivity ratio and contact angle, including attraction without coalescence (Mode I), attraction with coalescence (Mode II), coalescence with bubble entrapment (Mode III), coalescence followed by suspension (Mode IV) and suspension followed by coalescence (Mode V).

\subsubsection{Effect of permittivity ratio}
In this part, we mainly focus on the effect of permittivity ratios ($S$) on the interaction of a droplet pair with $\theta = 90^{\circ}$. As shown in Fig. \ref{picture-sp_90}, with the increase of $S$, the final state of the droplet pair changes from non-coalescence to coalescence, which indicates that the electrostatic interaction that promotes droplet coalescence becomes more dominant. Moreover, a larger $S$ brings a more significant deformation during coalescence, causing the upper portions of the droplets to contact first, which further leads to air  bubble entrapped [see Fig. \ref{picture-90_8}]. To understand this phenomenon more clearly, we present the distribution of the electrical stress ($\mathbf{F}_E$, directed from the droplet to the surrounding fluid) in Fig. \ref{picture-sp_lxt} where $t^*=675$. From this figure, one can see that for all cases, $\mathbf{F}_E$ only plays an important role in the regions near the interfaces of the droplets, resulting in horizontal deformation. However,  increasing $S$ weakens the lateral electrical stress, thereby reducing the horizontal deformation of droplets. Unlike the symmetric configurations in Figs. \ref{picture-sp_lxt_2} and \ref{picture-sp_lxt_3}, the droplets in Fig. \ref{picture-sp_lxt_1} experience asymmetric deformation. This asymmetry arises from the asymmetric distribution of $\mathbf{F}_E$ across the droplets. In addition to the distribution of electrical stress, the flow field around the droplets is also shown in Fig. \ref{picture-sp_lxt}, and four vortices form inside and outside the droplets. When $R>S$, the fluid flows from the equator toward the polar regions along the droplet interfaces, while if $R<S$, the flow direction reverses, moving from the poles to the equator, which is consistent with the vorticity orientation observed for the case of a single droplet \cite{Liu2019}.
\begin{figure}[h]
	\centering
	\includegraphics[width=0.5\textwidth]{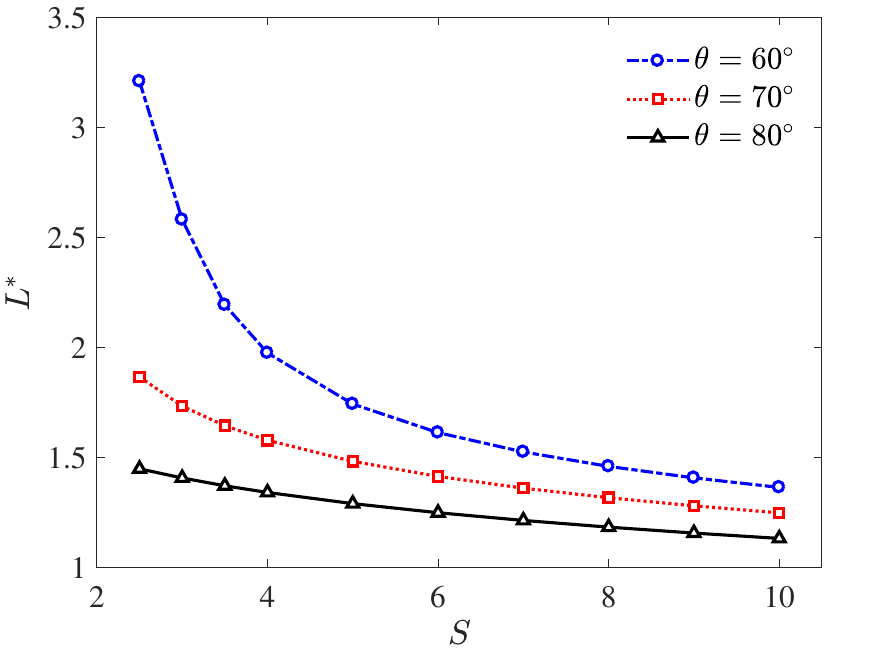}
	\vspace{-5pt}
	\caption{The spreading length $L^*$ after the coalescence of two droplets.}
	\label{picture-rhcd}
\end{figure}

Next, we conduct a quantitative study on the effect of $S$ on droplet coalescence, and focus on Model II. In particular, the normalized spreading length $L^*=L/(2 \sqrt{2} a)$ of post-coalescence hydrophilic droplets is measured under different values of permittivity ratio $S$. As seen from Fig. \ref{picture-rhcd}, the spreading length $L^*$ exhibits a monotonic decrease as $S$ increases, indicating that the axial elongation of the stable droplet along the electric field direction diminishes, which also agrees well with the theoretical predictions [Eqs. (\ref{Taylor}) and (\ref{Feng})]. Furthermore, due to the electrowetting of droplet, $L^*$ increases with increasing contact angle $\theta$, as seen from in Fig. \ref{picture-rhcd}.

\subsubsection{Effect of wettability}
\begin{figure}[h]
	\centering
	\includegraphics[width=0.5\textwidth]{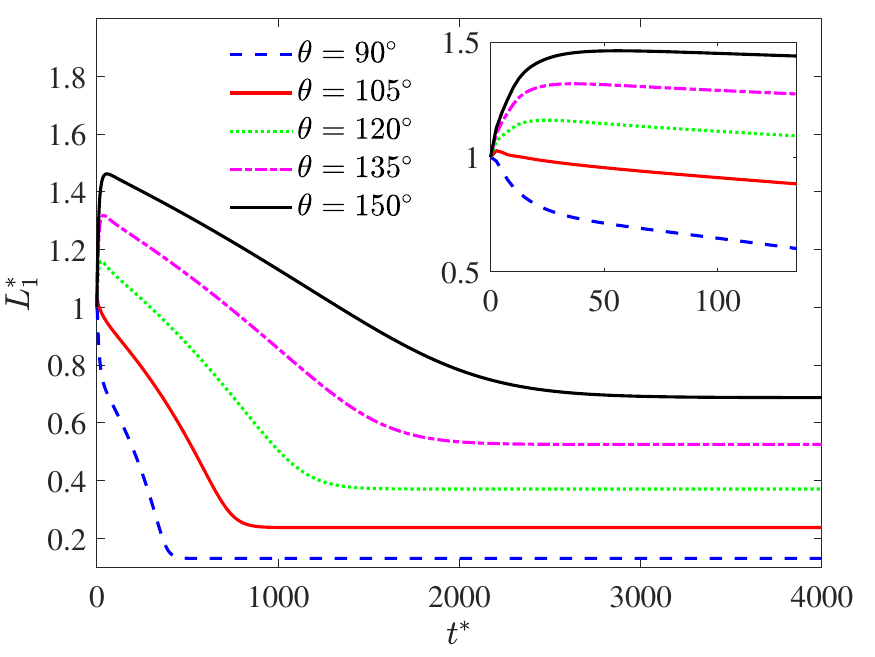}
	\vspace{-5pt}
	\caption{The evolutions of dimensionless inner-end distance $L_1^*$ at different contact angles.}
	\label{picture-jlbht}
\end{figure}

We further focus on the role of wettability in the coalescence process of droplets. As depicted in Fig. \ref{picture-xt1}, for the cases with small values of $S$ ($S\leq3.5$), an increase in contact angle suppresses droplet coalescence, and the system changes from a coalescence to a non-coalescence state. This is because the increase in contact angle reduces the lateral spreading of the droplets, thereby diminishing the likelihood of interfacial contact. To further investigate the influence of contact angle on the final state of the two droplets, we fix the permittivity ratio $S=2$ and plot the evolution of dimensionless inner-end distance ($L_1^*$) in time ($t^*$) in Fig. \ref{picture-jlbht}. As shown in this figure, the distance $L_1^*$ experiences an apparent increase during the initial stage, which is attributed to the shrinkage behavior of the droplet on a hydrophobic surface. Subsequently, due to the electrostatic interaction between the droplets, they gradually approach each other, resulting in a decrease of $L_1^*$. It should be noted that at $S=2$,  the relatively weak electrostatic interaction and a hydrophobic nature of the solid substrate prevent the coalescence of two droplets, eventually causing $L_1^*$ to remain constant. Furthermore, the results in Fig. \ref{picture-jlbht} also demonstrate that a larger contact angle yields a higher initial peak and slower decay rate, which indicates that the increase of the contact angle would suppress the droplet coalescence.
\begin{figure}[]
	\centering
	\subfigure[$S = 4$]{
		\begin{minipage}[b]{0.46\linewidth}
			\centering
			\hspace{-8.6mm}
			\includegraphics[width=3.0in]{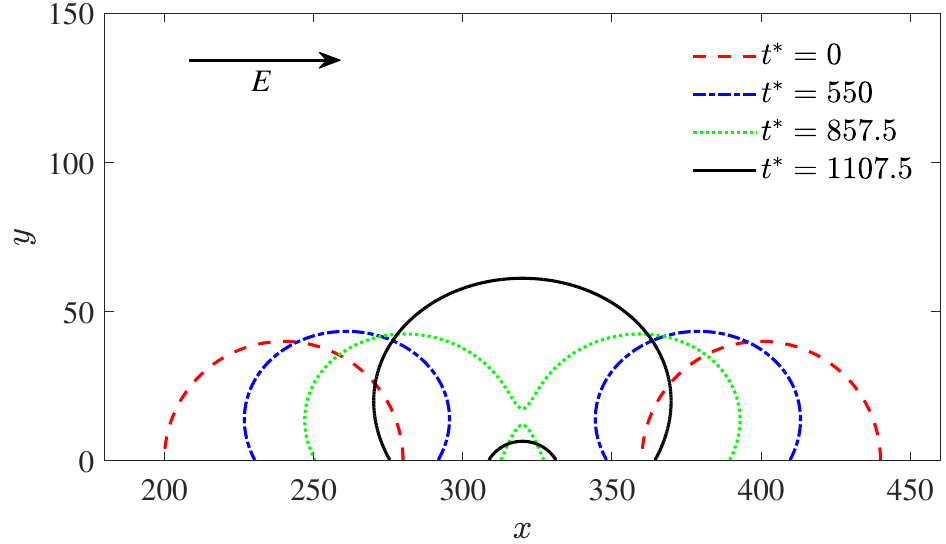}
		\end{minipage}
		\label{picture-120_4}}
	\subfigure[$S = 20$]{
		\begin{minipage}[b]{0.46\linewidth}
			\centering
			\hspace{-8.8mm}
			\includegraphics[width=3.0in]{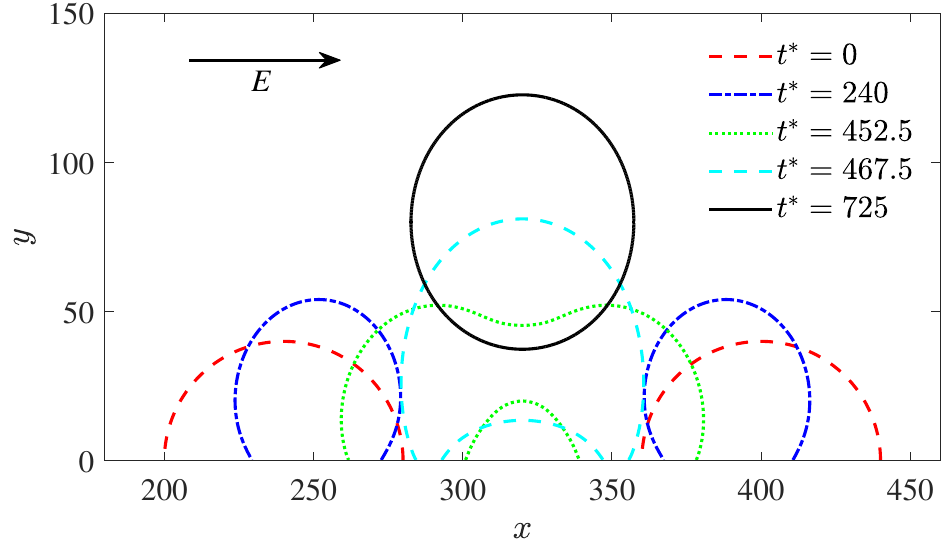}
		\end{minipage}
		\label{picture-120_20}}
	\caption{The interaction of the droplet pair at different permittivity ratios ($\theta=120^{\circ}$).}
\end{figure}
\begin{figure}[]
	\centering
	\subfigure[$S=4$]{
		\begin{minipage}[b]{0.46\linewidth}
			\centering
			\hspace{-7.6mm}
			\includegraphics[width=3.0in]{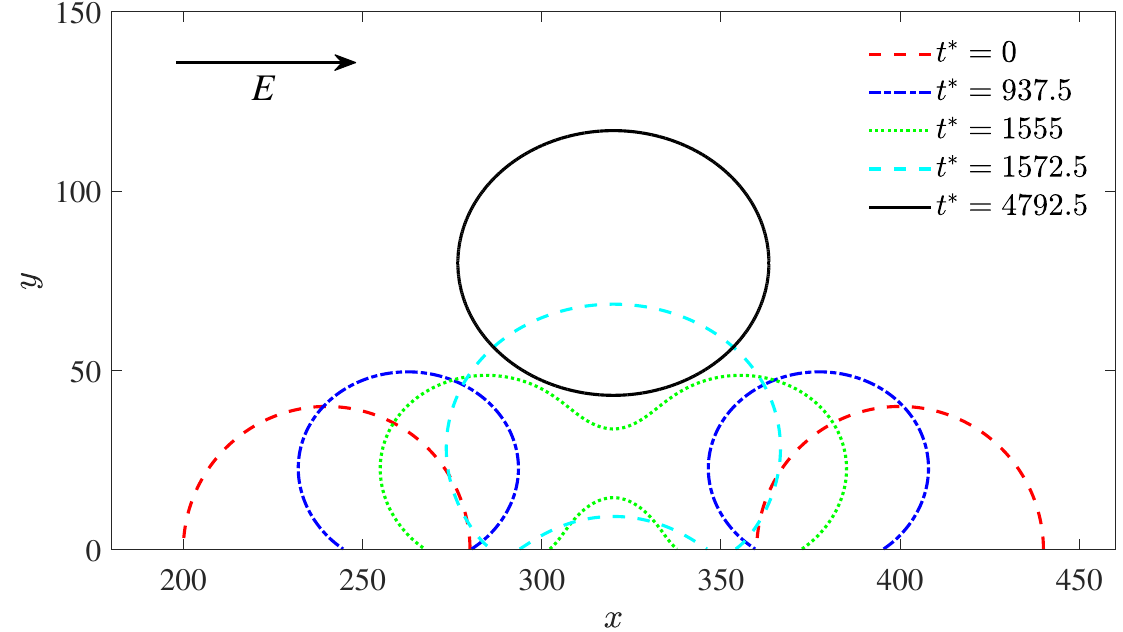}
		\end{minipage}
		\label{picture-150_4}}
	\subfigure[$S=20$]{
		\begin{minipage}[b]{0.46\linewidth}
			\centering
			\hspace{-7.9mm}
			\includegraphics[width=3.0in]{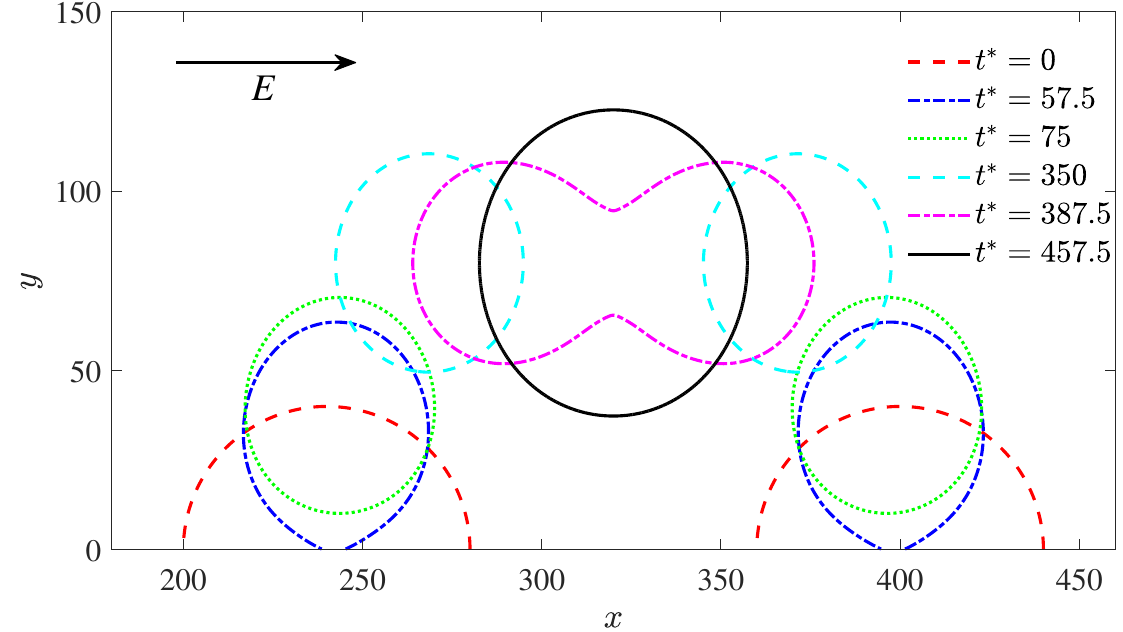}
		\end{minipage}
		\label{picture-150_20}}
	\caption{The interaction of the droplet pair at different permittivity ratios ($\theta=150^{\circ}$).}
\end{figure}
\begin{figure}[]
	\centering
	\subfigure[$t^*=467.5$]{
		\begin{minipage}[b]{0.475\linewidth}
			\centering
			\hspace{-1.4mm}
			\includegraphics[width=3.0in]{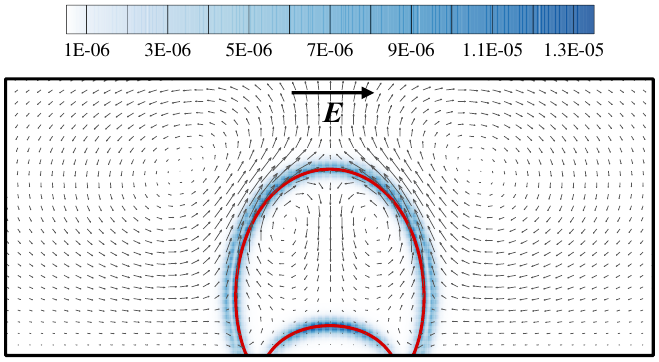}
		\end{minipage}
		\label{picture-sp_lxt_4}}
	\subfigure[$t^*=1572.5$]{
		\begin{minipage}[b]{0.475\linewidth}
			\centering
			\hspace{-1.4mm}
			\includegraphics[width=3.0in]{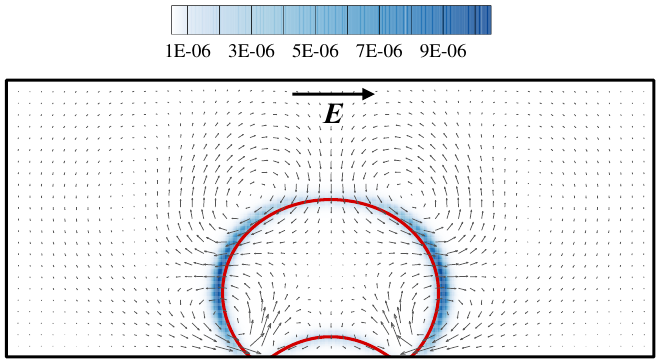}
		\end{minipage}
		\label{picture-sp_lxt_5}}
	
	\subfigure[$t^*=57.5$]{
		\begin{minipage}[b]{0.475\linewidth}
			\centering
			\hspace{-1mm}
			\includegraphics[width=3.0in]{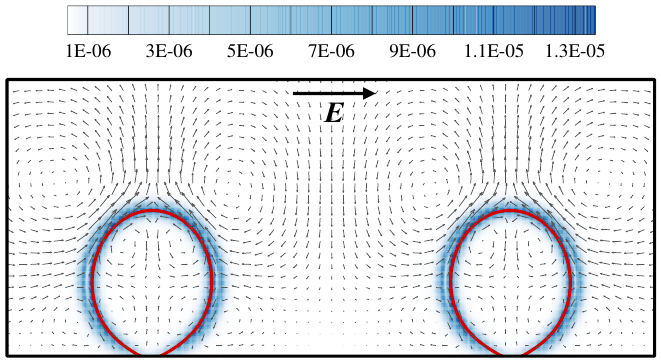}
		\end{minipage}
		\label{picture-sp_lxt_6}}
	\caption{The distribution diagram of the electrical stress $\mathbf{F}_E$ at different contact angles and permittivity ratios [(a) $(\theta,S)=(120^{\circ},20)$, (b) $(\theta,S)=(150^{\circ},4)$, (c) $(\theta,S)=(150^{\circ},20)$].}
	\label{picture-sp_lxt2}
\end{figure}

When $S$ is large ($S\geq4$), although the two droplets eventually coalesce, there are still some differences under specific contact angles. For instance, when $90^{\circ} \leq \theta \leq 130^{\circ}$, two droplets would coalesce, and two different modes [Mode III: coalescence with bubble entrapment in Fig. \ref{picture-120_4} and Mode IV: coalescence followed by suspension in Fig. \ref{picture-120_20}] are observed. Unlike the cases in Fig. \ref{picture-90_8}, the droplets in Fig. \ref{picture-120_4} are in a hydrophobic state, leading to non-bottom contact during coalescence and air bubble entrapment. When $S$ increase to 20, the pattern shifts from Mode III to Mode IV because the electric stress ($\mathbf{F}_E$, directed from the droplet to the surrounding fluid) concentrates at the droplet apex [see Fig. \ref{picture-sp_lxt_4}]. The deformation in the vertical direction results in significant bubble entrapment, and provides an upward lift, enabling the droplet to suspend after the coalescence. In addition, when $\theta \geq 140^{\circ}$, there are also two different patterns, the Mode IV [coalescence followed by suspension in Fig. \ref{picture-150_4}] and Mode V [suspension followed by coalescence in Fig. \ref{picture-150_20}]. Different from the case in Fig. \ref{picture-120_4}, the droplets in Fig. \ref{picture-150_4} are in a superhydrophobic state. Although $\mathbf {F}_E $ is mainly distributed on both sides of the droplet [see Fig. \ref{picture-sp_lxt_5}], the conversion between surface energy and kinetic energy during coalescence can detach the merged droplet from the substrate (coalescence-induced droplet jumping \cite{Liu2021}). Further increasing $S$ to 20 leads to the droplets subject to an upward electric stress $\mathbf{F}_E$ [see Fig. \ref{picture-sp_lxt_6}], directly detaching them from the substrate. Since $R<S$, both electrostatic and electrohydrodynamic interactions make the droplets approach continuously after detachment, leading to the coalescence [Fig. \ref{picture-150_20}].
\begin{figure}[]
	\centering
	\subfigure[Phase diagram II]{
		\begin{minipage}[b]{0.7\linewidth}
			\centering
			\hspace{-4.5mm}
			\includegraphics[width=4.5in]{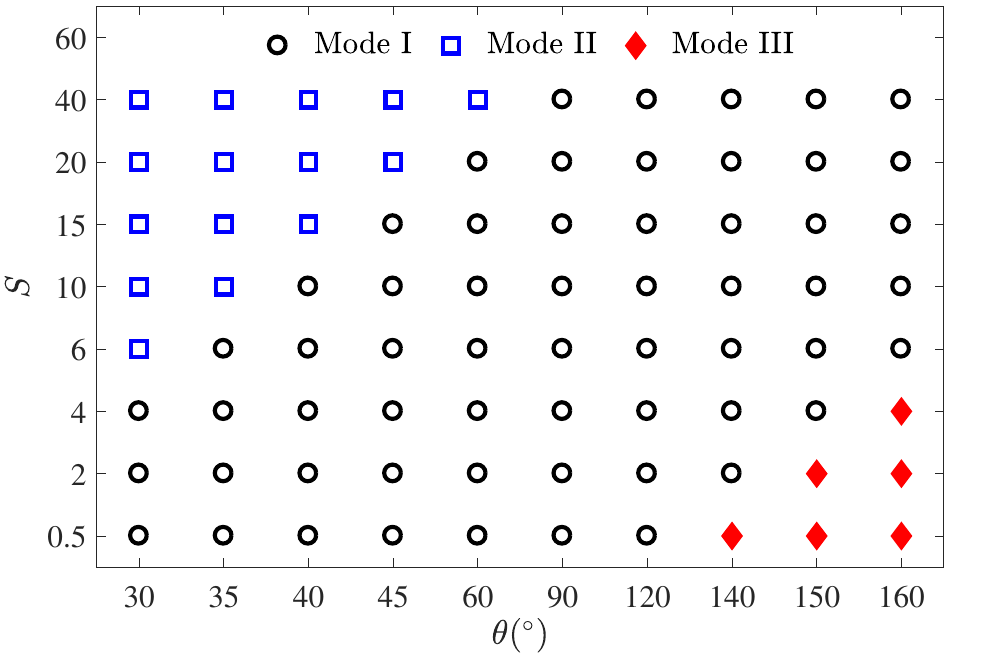}
		\end{minipage}
		\label{picture-xt2}}
	
	\subfigure[Modes I (left), II (middle) and III (right)]{
		\begin{minipage}[b]{1.0\linewidth}
			\centering
			\hspace{-7.5mm}
			\includegraphics[width=2.0in]{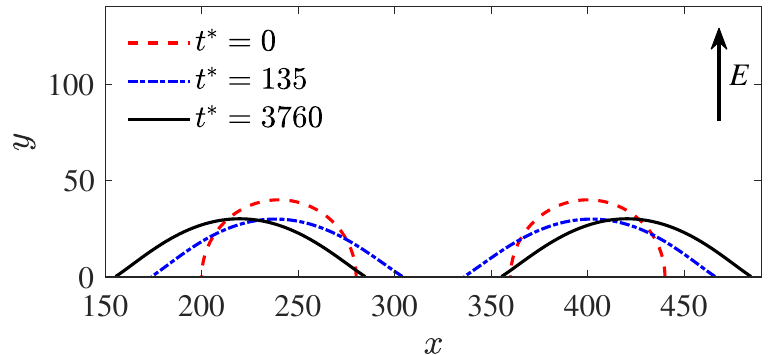}
			\includegraphics[width=2.0in]{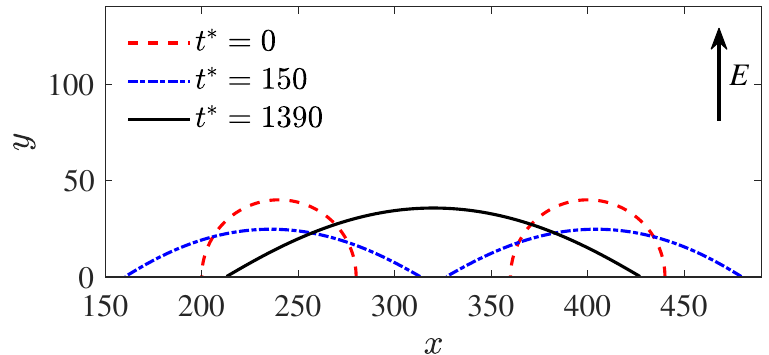}
			\includegraphics[width=2.0in]{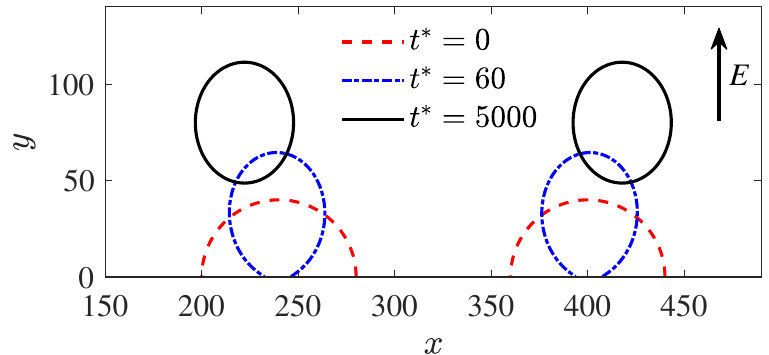}
		\end{minipage}
		\label{picture-mode2_1_2_3}}
	\caption{Electrohydrodynamics of the droplet pair at different contact angles ($\theta$) and permittivity ratios ($S$) [{\color{black}{$\circ$}}: non-coalescence, {\color{blue}{$\square$}}: coalescence, {\color{red}{$\blacklozenge$}}: suspension followed by repulsion].}
\end{figure}
\begin{figure}[h]
	\centering
	\subfigure[]
	{
		\begin{minipage}[b]{0.31\linewidth}
			\hspace{-2.5mm}
			\vspace{-0.3mm}
			\includegraphics[width=2.0in]{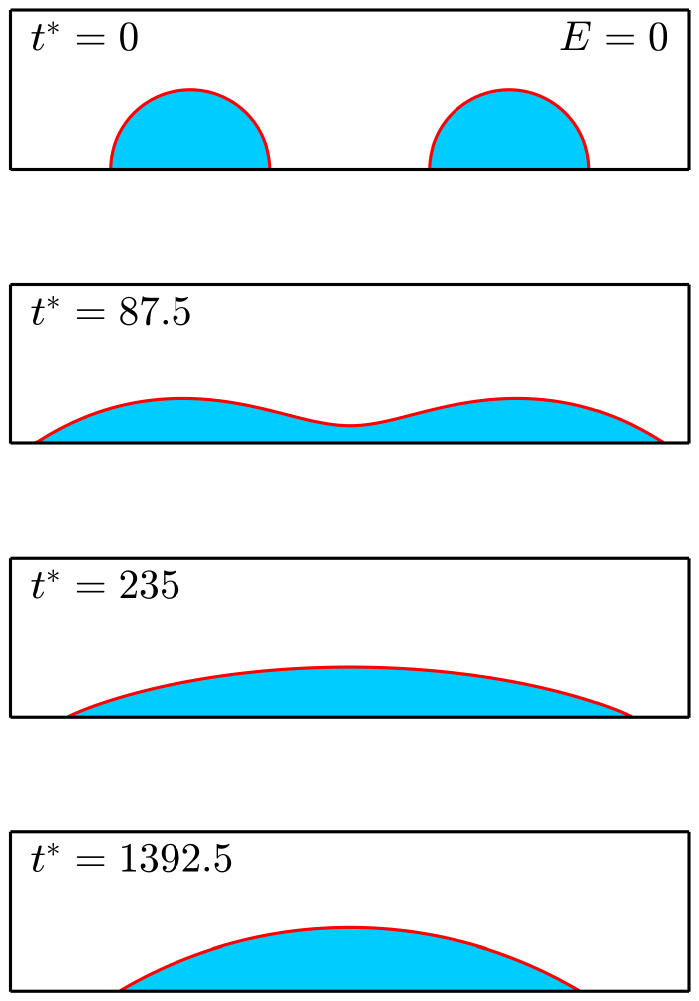}
		\end{minipage}
		\label{picture-30_0}
	}
	\subfigure[]
	{
		\begin{minipage}[b]{0.31\linewidth}
			\hspace{-2.5mm}
			\vspace{-0.3mm}
			\includegraphics[width=2.0in]{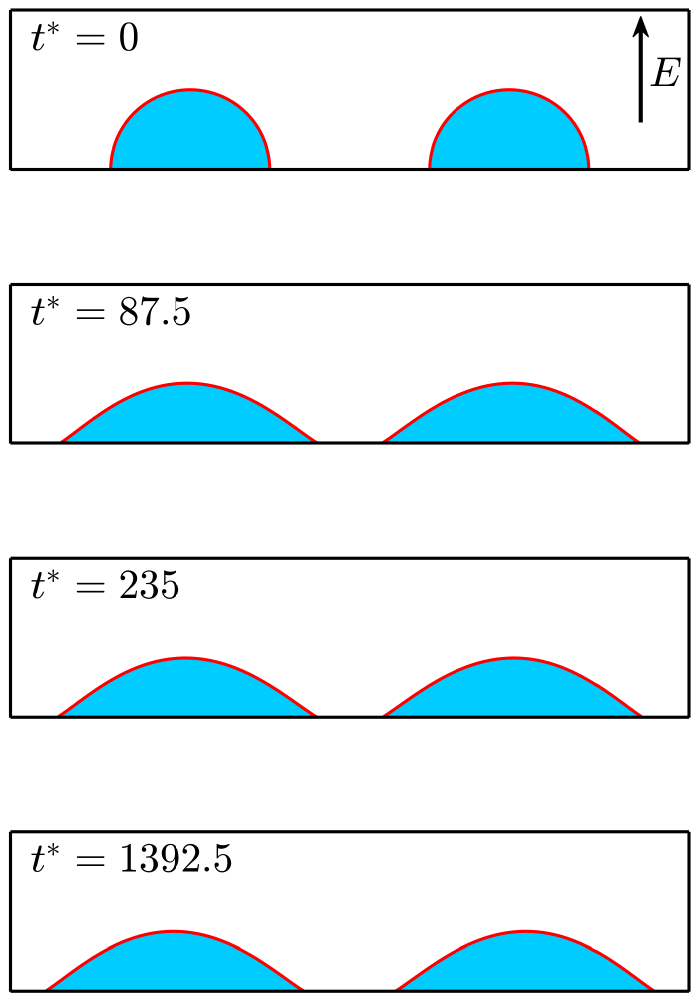}
		\end{minipage}
		\label{picture-30_0.5}
	}
	\subfigure[]
	{
		\begin{minipage}[b]{0.31\linewidth}
			\hspace{-2.5mm}
			\includegraphics[width=2.0in]{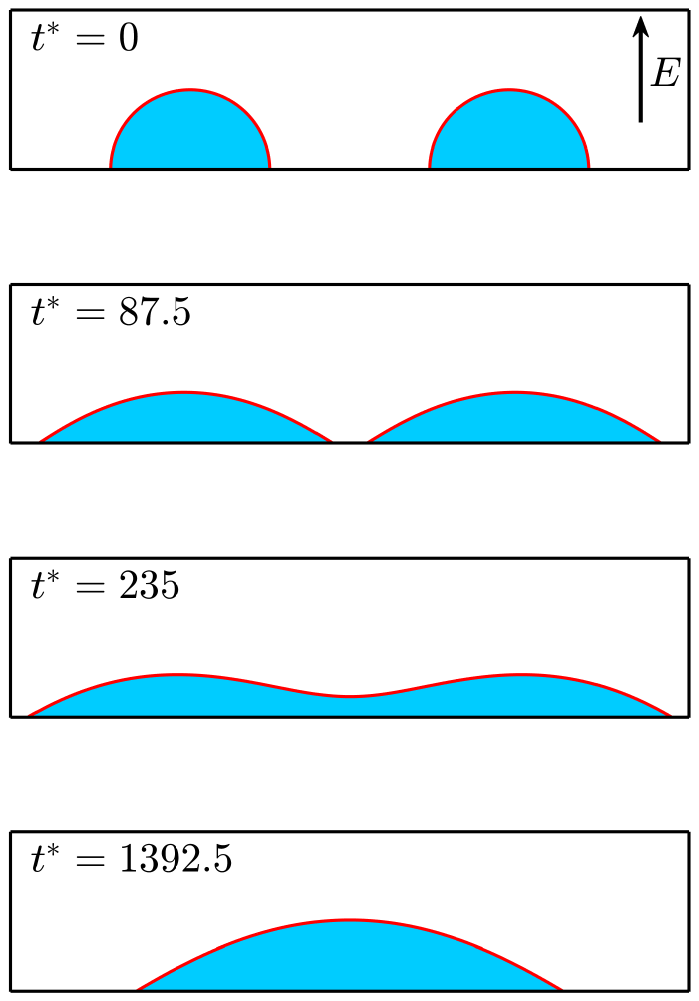}
		\end{minipage}
		\label{picture-30_6}
	}
	\caption{The evolutions of the droplet pair at different dimensionless times ($t^*$) [(a) the case without electric field, (b) $S=0.5$, (c) $S=6$].}
	\label{picture-sz_30}
\end{figure}
\begin{figure}[h]
	\centering
	\subfigure[]{
		\begin{minipage}[b]{0.475\linewidth}
			\centering
			\hspace{-2.5mm}
			\includegraphics[width=3.0in]{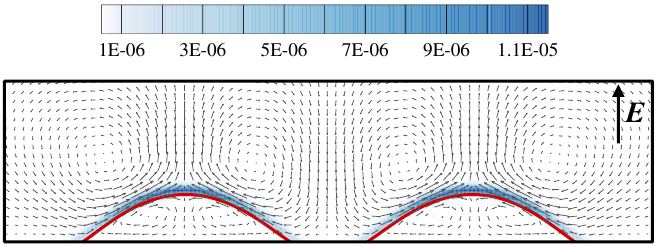}
		\end{minipage}
		\label{picture-sz_lxt_1}
	}
	\subfigure[]{
		\begin{minipage}[b]{0.475\linewidth}
			\centering
			\hspace{-2.5mm}
			\includegraphics[width=3.0in]{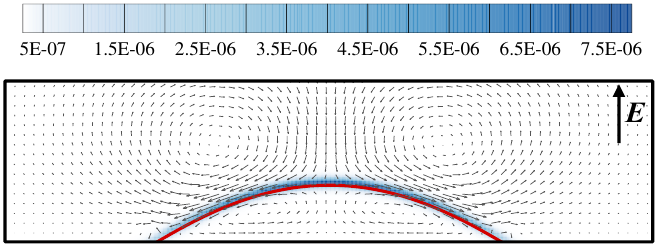}
		\end{minipage}
		\label{picture-sz_lxt_2}
	}
	\caption{The distributions of the electrical stress $\mathbf{F}_E$ at different permittivity ratios [(a) $S=0.5$, (b) $S=6$].}
	\label{picture-sz_lxt}
\end{figure}

\subsection{The droplet pair in a vertical external electric field}
We now investigate the effects of the permittivity ratio and wettability on the dynamic behavior of the droplet pair under a vertically upward electric field. Due to the change in the direction of the electric field, the Dirichlet boundary conditions with $\phi_0=0$ and $\phi_{\max}=\phi_0+E_0 L_y$ are imposed at the top and bottom boundaries. We perform some simulations on the electrohydrodynamics of the droplet pair, and present the phase diagram in Fig. \ref{picture-xt2} where three distinct modes are observed, including non-coalescence (Mode I), coalescence (Mode II) and suspension followed by repulsion (Mode III) [see Fig. \ref{picture-mode2_1_2_3}]. When the electric field is applied in vertical direction ($\alpha=90^{\circ}$), only the horizontal component $F_{h}$ of the electric force with a negative value exists [Eq. (\ref{dxhzy})], indicating a repulsive electrostatic interaction between the droplets.

\subsubsection{Effect of the permittivity ratio}
Similar to above discussion, we first consider the effect of permittivity ratio on the electrohydrodynamics of the droplet pair in Fig. \ref{picture-sz_30} where $\theta=30^\circ$. In the absence of an external electric field [see Fig. \ref{picture-30_0}], the two droplets contact and eventually coalesce. On the contrary, when an electric field is applied [see Fig. \ref{picture-30_0.5} where $S=0.5$], the droplets do not coalesce because the horizontal spreading length is reduced due to the repulsion between the droplets. When $S$ is increased to 6 [see Fig. \ref{picture-30_6}], the droplets contact and coalesce due to the significant deformation in the horizontal direction. To give a more detailed analysis of these phenomena, we present the distribution of electrical stress in Fig. \ref{picture-sz_lxt} where $t^*=1392.5$. From this figure, one can see that for both cases of $S=0.5$ and $S=6$, $\mathbf{F}_E$ is mainly distributed at the tops of the droplets, reducing the spreading length, compared to the case without electric field. We also note that the magnitude of $\mathbf{F}_E$ for the case $S=6$ is much weaker than that of the case $S=0.5$, resulting in an increase in spreading length, and consequently, the coalescence phenomenon can be observed. In addition, the vortex directions near the droplet interfaces in Fig. \ref{picture-sz_lxt} are also consistent with those of a single droplet.
\begin{figure}[h]
	\centering
	\includegraphics[width=0.5\textwidth]{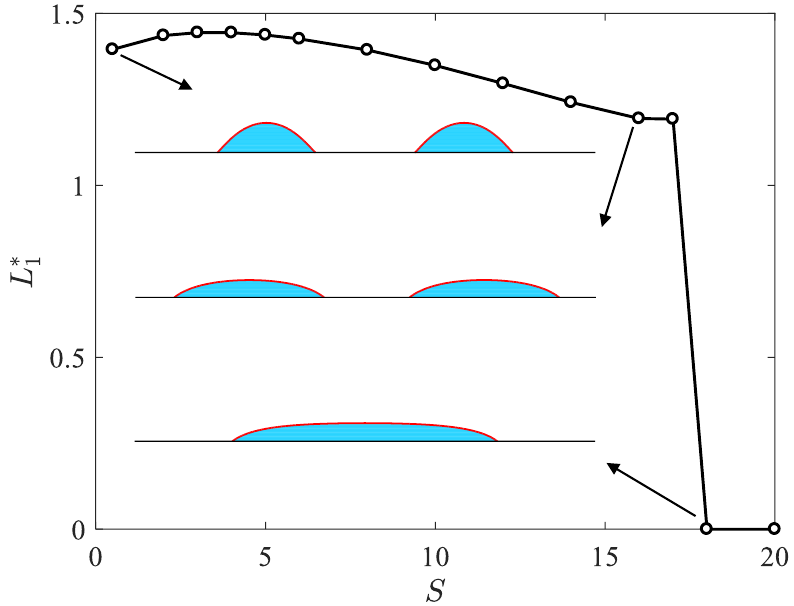}
	\vspace{-3mm}
	\caption{The dimensionless inner-end distance $L_1^*$ under different permittivity ratios ($S$).}
	\label{picture-45}
\end{figure}

To further investigate the influence of $S$ on the electrohydrodynamics of the droplet pair, we fix the contact angle $\theta=45^{\circ}$ and plot the dimensionless inner-end distance ($L_1^*$) in Fig. \ref{picture-45}. Generally, the electrostatic interaction is enhanced as $S$ increases, leading to an increase in $L_1^*$. However, when the permittivity ratio is larger than a critical value, the horizontal deformation of the droplets is increased due to the dominant hydrodynamic effect, resulting in a decrease in $L_1^*$. 

\subsubsection{Effect of wettability}
\begin{figure}[h]
	\centering
	\includegraphics[width=0.5\textwidth]{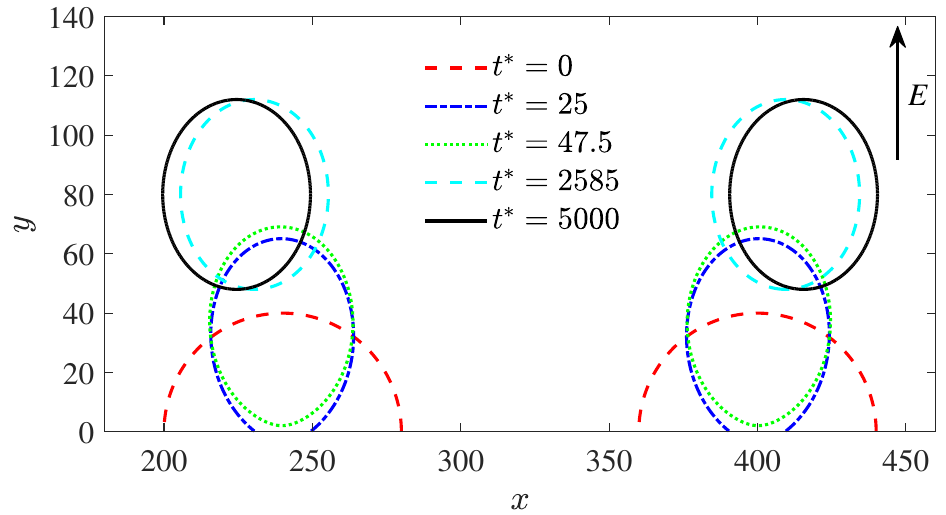}
	\vspace{-2mm}
	\caption{The interaction of the droplet pair under the condition of $\theta=150^{\circ}$ and $S=0.5$.}
	\label{picture-150_0.5}
\end{figure}

We now consider the effect of wettability on the electrohydrodynamics of the droplet pair. As shown in Fig. \ref{picture-xt2}, two droplets on a hydrophilic substrate are more likely to coalesce. This is because the droplets in a hydrophilic state would be prone to spread under the capillary interaction \cite{Pa1996}, making them easier to contact and coalesce. It is worth noting that when $\theta \geq 140^{\circ}$ and $S$ is small, the droplets exhibit an interesting phenomenon: they first suspend and then move apart, as illustrated in Fig. \ref{picture-150_0.5}. Similar to the results in Fig. \ref{picture-150_20}, the droplets in a superhydrophobic state can also detach directly from the solid substrate due to the upward electrical stress acting on their tops [see Fig. \ref{picture-150_0.5}], and then gradually move apart under the repulsive electrostatic interaction.

\begin{figure}[h]
	\centering
	\includegraphics[width=0.5\textwidth]{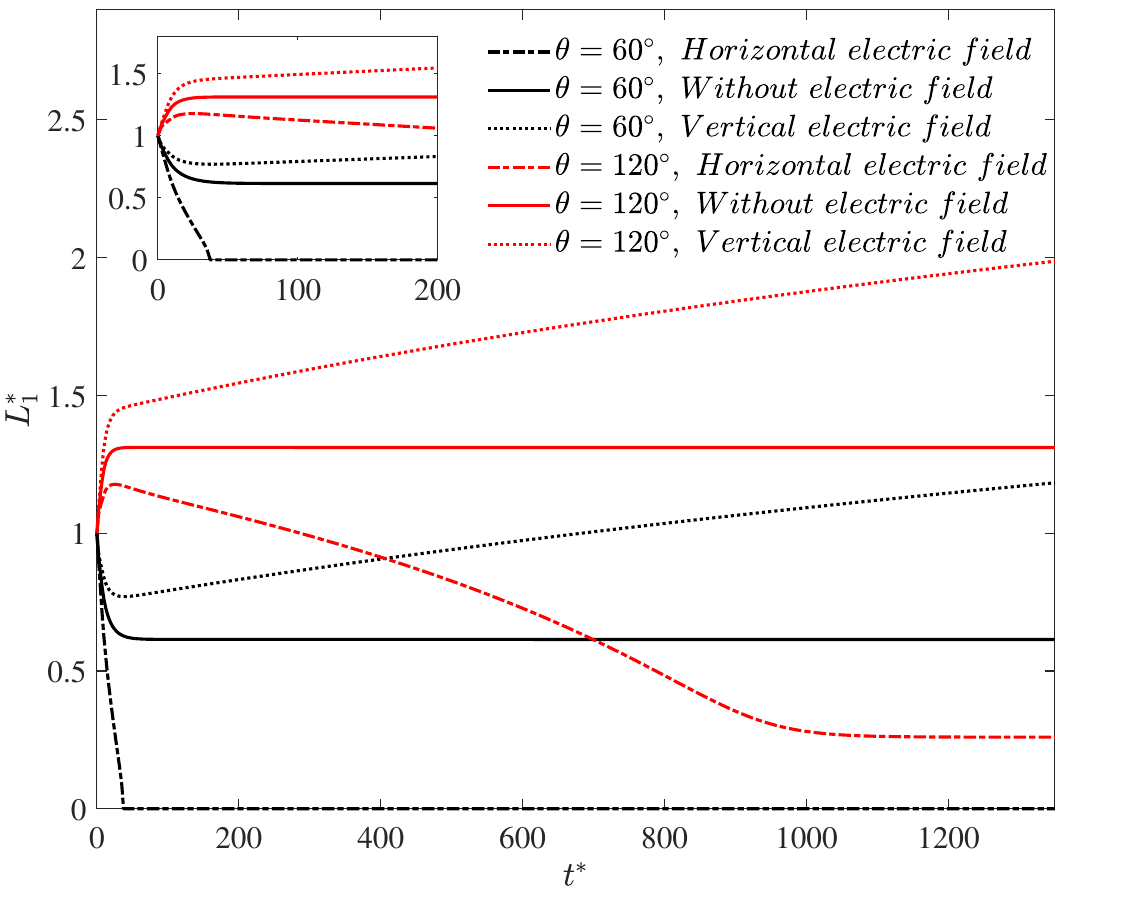}
	\vspace{-2mm}
	\caption{The evolutions of dimensionless inner-end distance $L_1^*$ under different electric field directions.}
	\label{picture-jddb}
\end{figure}

We also plot the evolutions of dimensionless inner-end distance $L_1^*$ under different electric field directions in Fig. \ref{picture-jddb} where $S = 3$, $\theta=60^{\circ}$ and $120^{\circ}$. As seen from this figure, regardless of whether the droplets are in a hydrophilic or hydrophobic state, the droplets under a horizontal electric field  eventually attract each other, while under a vertical electric field, they repel each other in the final stage. Furthermore, from Figs. \ref{picture-xt1} and \ref{picture-xt2}, one can observe that apart from the contact coalescence phenomenon caused by droplet spreading and the electrical stress, one can use the direction of the electric field to change the mode of the droplet pair. Additionally, through a comparison between the cases of $\theta=60^{\circ}$
and $\theta=120^{\circ}$, we can find that the droplets in a hydrophobic state are more difficult to coalesce when a vertical electric field is applied.

\section{Conclusions}\label{sec5}
In this paper, the electrohydrodynamics of a pair of leaky dielectric droplets on the solid substrate is studied by a consistent and conservative phase-field-based LBM. When a horizontal electric field is applied, there are five different modes of droplet pair, including attraction without coalescence (Mode I), attraction with coalescence (Mode II), coalescence with bubble entrapment (Mode III), coalescence followed by suspension (Mode IV) and suspension followed by coalescence (Mode V). For the first three modes, when the droplets are in a hydrophilic state, the spreading of the droplets promotes their coalescence. However, if the droplets are in a neutral or hydrophobic state, whether they can coalesce depends on the permittivity ratio $S$. Actually, the larger $S$ is, the easier they coalesce, this is because the electrostatic interaction  becomes more significant for  larger $S$. In addition, with increasing $S$,  the horizontal deformation of the droplets decreases, and the bubble is captured during the coalescence process. For Mode IV, there are two mechanisms for droplet suspension. On the one hand, the two droplets are separated from the solid substrate by an upward electrical stress after coalescence. On the other hand, for the droplet pair on a superhydrophobic substrate, surface energy is converted into kinetic energy during coalescence process, leading to suspension after coalescence. For Mode V, the droplets on a superhydrophobic substrate are subjected to an upward electrical stress ($\mathbf{F}_E$), and they first move away from the solid substrate, and then coalesce due to the electrostatic and electrically driven hydrodynamic interactions. 

When an electric field is applied in the vertical direction, the droplet pair exhibits three different modes, i.e., non-coalescence (Mode I), coalescence (Mode II), and suspension followed by repulsion (Mode III) in which the electrostatic force is repulsive. Compared to the coalesced droplets on the hydrophilic wall under no electric field, adjusting $S$ can prevent the droplets from coalescing. This is because the droplets are subjected to the upward electrical stress ($\mathbf{F}_E$), which reduces the spreading length, and the repulsive electrostatic force drives them apart. However, increasing $S$ induces horizontal deformation of the droplets, leading to coalescence. In addition, for the two droplets on a superhydrophobic substrate, the vertical electric field causes them move away after suspension due to the electrostatic and electrically driven hydrodynamic interactions, which differs from the phenomenon of coalescence after suspension under a horizontal electric field.

We hope that the research on the electrohydrodynamics of droplet pairs on a solid substrate provides guidance for controlling droplet dynamics through the application of an external electric field.

\section*{Acknowledgements}
The computation is completed in the HPC Platform of Huazhong University of Science and Technology. This work was financially supported by the National Natural Science Foundation of China (Grant No. 123B2018), the Postdoctoral Fellowship Program of CPSF (Grant No. GZB20250714), China Postdoctoral Science Foundation (Grant No. 2025M773077) and the Interdisciplinary Research Program of Hust (Grant No. 2024JCYJ001).

\normalem
\bibliography{Manuscript}

\end{document}